\documentclass[10pt]{article}

\usepackage{amsmath}
\DeclareMathOperator*{\Min}{min}
\DeclareMathOperator*{\Max}{max}
\usepackage{amssymb}

\usepackage{graphicx}

\usepackage{cite}

\usepackage{color} 
\usepackage{placeins}

\usepackage{setspace} 
\usepackage[labelfont=bf,font=footnotesize]{caption}
\usepackage{authblk}

\begin{document}
%
%\begin{flushleft}
%{\Large
%\textbf{Multichannel Compressive Sensing MRI Using Noiselet Encoding}
%}
%\\
%Kamlesh Pawar$^{1,2,3}$, 
%Gary Egan$^{4}$, 
%Jingxin Zhang$^{1,5,\ast}$
%\\
%\bf{1} Department of Electrical and Computer System Engineering, Monash University, Melbourne, Australia
%\\
%\bf{2} Indian Institute of Technology Bombay, Mumbai, India
%\\
%\bf{3} IITB Monash Research Academy, Mumbai, India
%\\
%\bf{4} Monash Biomedical Imaging, Monash University, Melbourne, Australia
%\\
%\bf{5} School of Software and Electrical Engineering, Swinburne University of Technology, Melbourne, Australia
%\\
%$\ast$ E-mail: jingxin.zhang@monash.edu
%\end{flushleft}

\title{Multichannel Compressive Sensing MRI Using Noiselet Encoding}
\date{}
\author[1,2,3]{Kamlesh Pawar}
\author[4]{Gary Egan}
\author[1,5,*]{Jingxin Zhang}
\affil[1]{Department of Electrical and Computer System Engineering, Monash University, Melbourne, Australia}
\affil[2]{Indian Institute of Technology Bombay, Mumbai, India}
\affil[3]{IITB Monash Research Academy, Mumbai, India}
\affil[4]{Monash Biomedical Imaging, Monash University, Melbourne, Australia}
\affil[5]{School of Software and Electrical Engineering, Swinburne University of Technology, Melbourne, Australia}
\affil[*] {Corresponding author: jingxin.zhang@monash.edu}

\maketitle

\begin{abstract}
The incoherence between measurement and sparsifying transform matrices \textcolor[rgb]{0.00,0.00,0.00}{and the restricted isometry property (RIP) of measurement matrix are two} of the key factors in determining the performance of compressive sensing (CS). In CS-MRI, the randomly under-sampled Fourier matrix is used as the measurement matrix and the wavelet transform is usually used as sparsifying transform matrix. However, the incoherence between the randomly under-sampled Fourier matrix and the wavelet matrix is not optimal, which can deteriorate the performance of CS-MRI. Using the mathematical result that noiselets are maximally incoherent with wavelets, this paper introduces the noiselet unitary bases as the measurement matrix \textcolor{black}{to improve the incoherence and RIP} in CS-MRI, and presents a method to design the pulse sequence for the noiselet encoding. This novel encoding scheme is combined with the multichannel compressive sensing (MCS) framework to take the advantage of multichannel data acquisition used in MRI scanners. \textcolor[rgb]{0.00,0.00,0.00}{An empirical RIP analysis is presented to compare the multichannel noiselet and multichannel Fourier measurement matrices in MCS.} Simulations are presented in the MCS framework to compare the performance of noiselet encoding reconstructions and Fourier encoding reconstructions at different acceleration factors. \textcolor[rgb]{0.00,0.00,0.00}{The comparisons indicate that multichannel noiselet measurement matrix has better RIP than that of its Fourier counterpart,} and that noiselet encoded MCS-MRI outperforms Fourier encoded MCS-MRI in preserving image resolution and can achieve higher acceleration factors. To demonstrate the feasibility of the proposed noiselet encoding scheme, \textcolor[rgb]{0.00,0.00,0.00}{two pulse sequences} with tailored spatially selective RF excitation pulses was designed and implemented on a 3T scanner to acquire the data in the noiselet domain from a phantom and a human brain.
\end{abstract}

\section{Introduction}
Magnetic resonance imaging (MRI) is a widely used imaging modality in clinical practice due to its ability to produce a good contrast between soft tissues and to image a slice at any orientation. The MRI scanner acquires data in a Fourier domain called k-space. MRI captures an image by scanning through the k-space on a Cartesian or non-Cartesian trajectory. This scanning process is time consuming and results in a long acquisition time and patient discomfort. Accelerating the data acquisition process is an active area of research in MRI, and Compressive Sensing (CS) is a promising solution that can improve the speed of data acquisition in MRI. CS \cite{donoho2006compressed,candes2006robust,candes2007sparsity,candes2005signal,candes2006near} is a technique that permits the faithful reconstruction of the signal of interest from the data acquired below the Nyquist sampling rate. MRI is a ideal system \cite{lustig2008compressed} for CS applications as it acquires image already in encoded form rather than in pixel domain. The application of CS in MRI was first described in \cite{lustig2007sparse}, where variable density random under-sampling of phase encodes was suggested as a sampling strategy. Parallel imaging techniques \cite{pruessmann1999sense,sodickson1997simultaneous,griswold2002generalized} have also been combined with CS in \cite{otazo2009distributed,liang2009accelerating,liang2008accelerating,otazo2010combination,lustig2010spirit,ji2008compressed} to further accelerate MRI scans, and CS-MRI has been applied for dynamic imaging, exploiting k-t space sparsity \cite{otazo2010combination,lustig2006kt,jung2009k,qiu2009real,jung2010motion}.

The theory of CS provides a solution to an ill-posed inverse problem by exploiting prior knowledge of signal sparsity or compressibility. This theory guarantees perfect reconstruction of the signal from the under-sampled data if certain conditions are satisfied \cite{donoho2006compressed,candes2006robust,candes2007sparsity,candes2005signal,candes2006near,candes2008restricted}: (i) sparsity or compressibility of the signal in the transform domain; (ii) Restricted Isometry Property (RIP) of the measurement matrix or incoherence between the measurement and sparsifying transform matrices; and (iii) a non-linear reconstruction algorithm that promotes sparse representation of the image and enforces data consistency of  reconstruction with the acquired data. 

The sparsity or compressibility condition is satisfied by MR images as they are known to be sparse or compressible in the wavelet domain and the finite difference domain \cite{lustig2007sparse,otazo2009distributed,liang2009accelerating,liang2008accelerating}. However, the RIP is difficult to verify for a given deterministic measurement matrix since it is computationally NP hard \cite{tillmann2012computational}. An empirical solution to this problem in the CS literature is to use random measurement matrices. A randomly sampled frequency domain data can capture pertinent information from a sparse signal with fewer measurements and allows accurate reconstruction of the signal by the convex $l_1$ optimization program. This property was first proved mathematically for Gaussian matrices \cite{candes2006near,candes2005decoding} and has recently been extended to a wide class of random matrices \cite{bayati2012universality}. Based on this property, \cite{haldar2011compressed} proposed using spatially selective RF pulses to implement random encoding along the phase encode direction, with the entries of the random measurement matrix drawn from Gaussian distribution. This random encoding scheme attempts to approximate the sufficient conditions for perfect CS reconstruction, but as described in \cite{haldar2011compressed}, this measurement matrix is not unitary and results in noise amplification even after taking all the required measurements. Another problem with random encoding is computational complexity. Dense random matrices consume large amounts of memory and require computationally expensive matrix multiplications in CS-reconstruction \cite{candes2007sparsity,Blanchard2013Towards}. This problem is partially alleviated in \cite{haldar2011compressed} by using fast Fourier transforms of the matrix multiplications, but still requires more memory and computations than those of structured/unitary measurement matrices.

MRI uses the Fourier basis to encode the excited region of interest. The Fourier measurement matrix is weakly incoherent with the wavelet sparsifying transform matrix, thus is sub-optimal for CS-MRI \cite{puy2012spread}. The incoherence is essentially a measure of the spread of sparse signal energy in the measurement domain \cite{candes2007sparsity}. Various attempts have been made in \cite{haldar2011compressed,puy2012spread,liang2009toeplitz,sebert2008compressed,wang2009toeplitz} to spread the energy of the MR signal in the measurement domain. In \cite{puy2012spread,qu2013spread}, the spread spectrum technique was presented which convolves the k-space with the Fourier transform of a chirp function to spread the energy of the MR signal in the measurement domain. The chirp modulation is implemented through the use of second order shim coils. In \cite{liang2009toeplitz,sebert2008compressed,wang2009toeplitz,liu2012compressive,wang2012three, Leslie2012three}, other non-Fourier encoding strategies were described for compressive sensing that aims to spread the energy of the MR signal in the measurement domain. While these encoding strategies can spread the signal energy to some extent, none of them has the theoretically proven maximal incoherence for the complete spread of the signal energy.

Noiselet bases \cite{coifman2001noiselets,tuma2009incoherence} are known to completely spread out the energy of the signal in the measurement domain, which is a desired property in CS. Noiselets are also known to be maximally incoherent with Haar wavelets that makes them the best suited bases for CS. Further, noiselet matrices are complex valued, symmetric and unitary, which simplifies the implementation of image reconstruction program in CS-MRI. \textcolor[rgb]{0.00,0.00,0.00}{In the simulation study of \cite{datta2013stability}, it is found that the noiselet measurement matrix outperforms the chirp modulation measurement matrix when the noise level is high. Also, as shown in Section III of this paper, the multichannel noiselet measurement matrix exhibits much better RIP than that of its Fourier counterpart.} In order to take the advantage of maximal incoherence \textcolor[rgb]{0.00,0.00,0.00}{and better RIP} provided by noiselet measurement matrix, we have investigated the use of noiselet encoding for CS-MRI.

In order to take the advantage of multiple measurements provided by an MR scanner through the use of multiple channels, a Multichannel Compressive Sensing (MCS) framework is proposed in \cite{otazo2009distributed} for CS reconstructions. The MCS framework simultaneously uses data from the multiple channels to reconstruct the desired image instead of reconstructing separate images from each channel, resulting in higher acceleration factors and improved image quality. Therefore, in this paper we describe the theory and implementation details of using noiselet bases as the measurement matrix in MCS-MRI. \textcolor[rgb]{0.00,0.00,0.00}{Considering the lack of analysis and sufficient understanding of MCS-MRI in the literature, we also present an empirical RIP analysis of the multichannel noiselet measurement matrix in comparison with its Fourier counterpart. The results indicate that the multichannel noiselet measurement matrix outperforms its Fourier counterpart,} and that noiselet encoding outperforms Fourier encoding in preserving image resolution for the same acceleration factors, and can achieve higher acceleration factors than the Fourier encoding scheme for the desired image quality and resolution.

The paper is organized as follows. In section II, we describe the background of CS, sufficient conditions for CS and develops a model for MCS-MRI reconstruction. In section III, we describe the noiselet basis function, its properties and our motivation for using noiselets in MRI. A pulse sequence design to implement the proposed noiselet encoding scheme is also described in this section, followed by an empirical RIP analysis of the multichannel noiselet measurement matrix in comparison with its Fourier counterpart. In section IV, simulation studies comparing the performance of noiselet encoded and Fourier encoded MCS-MRI for different acceleration factors are demonstrated on a brain image. The effect of the number of channels and level of noise on the reconstruction is also evaluated for both the encoding schemes. In section V, we demonstrate the feasibility of the proposed encoding scheme by acquiring noiselet encoded data from a phantom and a human brain. Retrospective under-sampling is performed on the acquired noiselet encoded and the Fourier encoded data to simulate accelerated acquisition. The nonlinear conjugate method \cite{lustig2008compressed} with wavelet and total variation (TV) penalties is used to solve the minimization program for MCS-MRI. In section VI, we discuss the findings, limitation and further extension of the technique.

\section{Compressive Sensing}
Compressive sensing is a mathematical theory describing how a sparse signals can be faithfully recovered after sampling projections well below the Nyquist sampling rate. Consider a signal $x$ in the $n$ dimensional complex space $\mathbb{C}^n$ that can be sparsely represented in $\Psi$ domain as $s=\Psi x$, where $\Psi$ is the $n\times n$ sparsifying transform matrix. The signal $s$ is $K$-sparse, that is, only the $K$ coefficients in $s$ are non-zero. A measurement system measures signal $y$ in $m$ dimensional space by taking only $m$ projections of the signal $x$ as
\begin{equation}
\label{eq:acq}
\begin{split}
y = \Phi x  %\hspace{25mm} \\
\end{split}
\end{equation}
where $y \in \mathbb{C}^m$, $K < m \ll n$ and $\Phi$ is the $m\times n$ measurement matrix. In MRI, $\Phi$ is usually a (partially) randomly under-sampled discrete Fourier transform matrix. Equation \eqref{eq:acq} can be further expressed as
\begin{equation}
y = \Phi \Psi^* s, \quad x = \Psi^* s
\end{equation}
where $^*$ represents the conjugate transpose operation and the signal $x$ is sparse in the $\Psi$ domain. MR images can be sparsely represented in the wavelet domain using the wavelet transform matrix. Given the measurement $y$ and the matrices $\Phi$ and $\Psi$, there exist many solutions satisfying \eqref{eq:acq} and recovering $x$ becomes an ill posed problem. The CS theory provides a unique solution to the ill-posed problem by solving the following optimization program:
\begin{equation}
\label{opt}
\Min_{\hat{x}} \hspace{3mm}\|\Psi\hat{x}\|_{l_1} \hspace{5mm} s.\hspace{1mm}t. \hspace{5mm} y =\Phi \hat{x}
\end{equation}
where $\|{x}\|_{l_1}:=\sum_{i} |x_i|$ is the $l_1$ norm of $x$, with $x_i$ the $i$th element of $x$. Exact reconstruction of the signal $x$ is achievable if certain mathematical conditions hold.

\textcolor[rgb]{0.00,0.00,0.00}{\subsection{Restricted Isometry and Incoherence in Compressive Sensing}
An important sufficient condition for exact reconstruction of $x$ is the so called restricted isometry property (RIP) \cite{candes2006robust,candes2008restricted,candes2005decoding}. For a normalized measurement matrix $\Phi$ with unit column norms, the RIP is given as
\begin{equation}
\label{eq:RIP}
(1-\delta_K) \|x\|_{l_2}^2 \leq  \|\Phi x\|_{l_2}^2\leq (1+\delta_K) \|x\|_{l_2}^2
\end{equation}
where $\delta_K \in (0,1)$ is called the RIP constant. The RIP (\ref{eq:RIP}) is equivalent  to \cite{candes2006robust,ni2011efficient,AHSC09}
\begin{equation}
\label{sigma}
(1-\delta_K) \leq \sigma^2_{min}[\Phi_{sub}(K)] \leq \sigma_{max}^2[\Phi_{sub}(K)] \leq (1+\delta_K),
\end{equation}
%which in turn implies
%\begin{equation}
%\label{sigma_1}
%1 \leq \frac{\sigma^2_{max}[\Phi_{sub}(K)]}{\sigma^2_{min}[\Phi_{sub}(K)]} \leq \frac{(1+\delta_K)}{(1-\delta_K)}.
%\end{equation} and (\ref{sigma_1})
where $\Phi_{sub}(K)$ is the $m \times K$ submatrix formed from $K$ distinct columns of $\Phi$,  and $ \sigma_{min}[\Phi_{sub}(K)]$ and  $\sigma_{max}[\Phi_{sub}(K)]$ are  the minimum and maximum singular values\footnote{\textcolor[rgb]{0.00,0.00,0.00}{In \cite{candes2006robust,ni2011efficient,AHSC09}, the eigenvalue $\lambda[\Phi_{sub}(K)^*\Phi_{sub}(K)]$ is used for (\ref{sigma}), where $\lambda[\Phi_{sub}(K)^*\Phi_{sub}(K)] = \sigma^2[\Phi_{sub}(K)]$.}} of $\Phi_{sub}(K)$, respectively.
%and the ratio $\frac{\sigma^2_{max}[\Phi_{sub}(K)]}{\sigma^2_{min}[\Phi_{sub}(K)]}$ $=:R$ is the squared condition number of $\Phi_{sub}(K)$. and (\ref{sigma_1})
The RIP constant $\delta_K$ is the smallest constant that satisfies the inequalities (\ref{sigma}) for every $m \times K$ sized submatrix of $\Phi$, and it is essentially a bound on the distance between unity and the singular values of all $\Phi_{sub}(K)$s. It is shown in \cite{candes2005decoding} that if $\delta_{2K} < 1$, then a $K$-sparse signal $x$ can be exactly reconstructed from the measurements of $\Phi$.}

\textcolor[rgb]{0.00,0.00,0.00}{While $\delta_K \in (0, 1)$ renders the exact reconstruction of $x$, the value of $\delta_K$ determines the stability of reconstruction. In the presence of measurement noise $\epsilon$, $y = \Phi x + \epsilon$ and the reconstructed signal $\hat{x}$ satisfies (Section 5.2 \cite{elad2010sparse})
\begin{equation}
\label{d_error}
\|x-\hat{x}\|^2_{l_2} \leq \frac{4 \epsilon^2}{1- \delta_{2K}}.
\end{equation}
Thus, the smaller the $\delta_K$, the smaller the reconstruction error, and vice versa. Since measurement noise always exists in practice, the value of $\delta_K$ is an important performance measure of a measurement matrix $\Phi$ for both the reconstruct-ability and reconstruction error. However, the computation of $\delta_K$ for a given $\Phi$ is NP hard and hence intractable. Since the RIP constant $\delta_K$ is essentially a bound on the distance between 1 and the singular values of all $\Phi_{sub}(K)$s, the value of $\delta_K$ can be assessed by the distances from 1 to the $\sigma_{min}[\Phi_{sub}(K)]$s and $\sigma_{max}[\Phi_{sub}(K)]$s. The smaller the distance, the smaller the $\delta_K$ and hence the better performance of $\Phi$. Since (\ref{sigma}) must hold for all the $m \times K$ submatrices of $\Phi$, the statistics of $\sigma_{min}[\Phi_{sub}(K)]$ and $\sigma_{max}[\Phi_{sub}(K)]$ over randomly sampled $\Phi_{sub}(K)$s are used in \cite{candes2006near,ni2011efficient,AHSC09} to assess the RIP performance of a given measurement matrix $\Phi$. This method is also adopted in this paper.}

\textcolor[rgb]{0.00,0.00,0.00}{Another important sufficient condition for exact reconstruction of $x$ is the incoherence \cite{candes2007sparsity,tuma2009incoherence}. For a pair of measurement matrix $\Phi$ and sparsifying transform matrix $\Psi$, satisfying $\Phi^*\Phi = nI$ and $\Psi^*\Psi = I$, their incoherence is defined as
\begin{equation}
\mu(\Phi,\Psi) = \Max_{k,j} |\left\langle \Phi_k,\Psi_j \right\rangle |
\end{equation}
where $\Phi_k$ and $\Psi_j$ are respectively the $k$th and $j$th columns of $\Phi$ and $\Psi$, and $\mu(\Phi,\Psi) \in [1,\sqrt{n}]$. The value $\mu(\Phi,\Psi) = 1$ is termed as maximal incoherence. As shown in \cite{candes2007sparsity}, if $m \geq C \cdot \mu^2(\Phi,\Psi)\cdot K \cdot log(n)$, where $C$ is a small constant, then a $K$-sparse signal $x$ can be exactly reconstructed. Thus, $\mu(\Phi,\Psi)$ determines the minimum number of measurements needed for exact reconstruction of $x$. The smaller the $\mu(\Phi,\Psi) $, the smaller the $m$ (the fewer the measurements) needed for exact reconstruction of $x$.
}

\textcolor[rgb]{0.00,0.00,0.00}{It is important to note that both the RIP and the incoherence are sufficient conditions on the measurement matrix. So they are parallel and either or both of them can be used to design, analyze and assess the measurement matrix for exact reconstruction of $x$.}

\subsection{Multichannel Compressive Sensing}
\textcolor[rgb]{0.00,0.00,0.00}{MRI systems acquire multiple measurements of the desired signal through multiple channels. Given the multiple channels, the data acquisition process can be modeled as
%\begin{equation}
%\label{eq:acq_pmri}
$y_i = \Phi\Gamma_i x = \Phi\Gamma_i \Psi^* s, i = 1,2,\cdots, L,$
%\end{equation}
where $\Gamma_i = diag[\gamma_{ij}]_{j=1,2, \cdots, n}$ is the complex valued sensitivity map matrix of the $i$th receive channel, with $\gamma_{ij}$ being the sensitivity of the $i$th channel at the $j$th pixel of the vectorized image, $L$ is the number of receive channels,  $y_i$ is the data acquired from the $i$th receive coil and $s = \Psi x$.
In matrix form, the $y_i$'s can be written as
\begin{equation}
\label{eq:acq_pmri}
Y:=
\begin{bmatrix}
y_1\\
y_2\\
\vdots \\
y_L
\end{bmatrix}
=
\begin{bmatrix}
\Phi\Gamma_1\\
\Phi\Gamma_2\\
\vdots \\
\Phi\Gamma_L
\end{bmatrix}
x
=:
Ex
= E \Psi^*s.
\end{equation}
As seen from above, with $L$ receive channels, the measurement matrix for $x$ becomes $E$, which has a column of $L$ measurement matrices $\Phi \Gamma_i$'s and dimension $Lm \times n$. It is important to note that  $\Gamma_i$s are complex valued and $\Gamma_i \neq \Gamma_j, i \neq j$, in general. Hence, $\Phi \Gamma_i \neq \Phi \Gamma_j$ for $i \neq j$ and they can be independent of each other, depending on the specific values of $\Gamma_i$ and $\Gamma_j$. As a result, the multichannel measurement matrix $E$ provides more independent measurements than that of the single channel $\Phi$,  which may reduce the number of measurements, $m$, needed at each channel for exact reconstruction of $x$. This reasoning is confirmed by the empirical analysis of $E$ in Section III.}

\textcolor[rgb]{0.00,0.00,0.00}{In light of the above discussion, the following MCS optimization is considered for reconstructing the desired image $x$ from the multichannel measurements of  MRI.
\begin{equation}
\label{eq:min_cs}
\Min_{\hat{x}} \hspace{3mm} \|\Psi \hat{x}\|_{l_1} \hspace{3mm} s.t. \hspace{3mm} \|Y-E \hat{x}\|_{l_2} \leq \epsilon
\end{equation}
where $\Psi$ is the wavelet transform operator and $\epsilon$ determines the allowed noise level in the reconstructed image.}

\textcolor{black} {MR images are also known to be sparse in the total variation (TV) domain. It is demonstrated in \cite{ma2008efficient} that the TV penalty is critical to the performance of CS-MRI, and that MR images can be recovered more efficiently with the use of TV penalty together with the wavelet penalty. Therefore, most of the CS-MRI work\cite{lustig2008compressed,lustig2007sparse,liang2009accelerating,liang2008accelerating,haldar2011compressed,liang2009toeplitz,wang2009toeplitz,wang2012three,Leslie2012three} has used both TV and wavelet penalties for better reconstruction performance. To be consistent with this common practice in CS-MRI, the TV penalty is included in the objective function \eqref{eq:min_cs} for MCS-MRI reconstruction, together with the wavelet penalty. Equation  \eqref{eq:min_cs} is a constrained optimization problem which is computationally intensive to solve. To relax the problem, \eqref{eq:min_cs} is converted to the unconstrained optimization problem with the inclusion of TV penalty.
\begin{equation}
\label{eq:min_tv_wav}
\Min_{\hat{x}} \hspace{3mm} \lambda_1\|\Psi \hat{x}\|_{l_1} + \lambda_2\| TV \hat{x}\|_{l_1} + \|Y-E \hat{x}\|_{l_2}^2
\end{equation}
where $TV$ is a 2D total variation operator and $\lambda_1$, $\lambda_2$ are regularization parameters for wavelet and $TV$ penalties, respectively.}

\textcolor[rgb]{0.00,0.00,0.00}{Daubechies-4 (db-4) wavelet is usually used in CS-MRI because of its superior performance in sparsifying the MR images. To be consistent with this fact and fair in comparison with the existing CS-MRI results, the unconstrained objective function (\ref{eq:min_tv_wav}) with the $\Psi$ of the db-4 wavelet operator will be used throughout all the simulations and reconstructions in this work.}

\section{Noiselet Encoding in CS-MRI}
\subsection{Noiselets}
Noiselets are functions which are noise-like in the sense that they are totally incompressible by orthogonal wavelet packet methods \cite{coifman2001noiselets,tuma2009incoherence}. Noiselet basis functions are constructed similar to the wavelet basis functions, through a multi-scale iteration of the mother bases function but with a twist. As wavelets are constructed by translates and dilates of the mother wavelet function, noiselets are constructed by twisting the translates and dilates \cite{candes2007sparsity}. The mother bases function $\chi(x)$ can be defined as
\[
  \chi(x) = \left\{
  \begin{array}{l l}
    1 & \quad \text{$x \in$  [0,1)}\\
    0 & \quad \text{otherwise}\\
  \end{array} \right.
\]
The family of noiselet basis functions are generated in the interval $[0,1)$ as
\begin{equation}
\begin{split}
f_1(x) &=\chi(x)\\
f_{2n}(x) &=(1-i)f_{n}(2x) + (1+i)f_{n}(2x-1) \\
f_{2n+1}(x) &=(1+i)f_{n}(2x) + (1-i)f_{n}(2x-1) \\
\end{split}
\end{equation}
where $i = \sqrt{-1}$ and $f_{2^n},\hdots\hdots, f_{2^n+1}$ form the unitary basis for the vector space $V_n$. An example of a 4$\times$4 noiselet transform matrix is given below.
\begin{equation}
\label{eq:nmat}
\frac{1}{2}\begin{bmatrix}
0-1i & 1+0i & 1-0i & 0+1i\\
1+0i & 0+1i & 0-1i & 1-0i  \\
1-0i & 0-1i & 0+1i & 1+0i \\
0+1i & 1-0i & 1+0i & 0-1i
\end{bmatrix}
\end{equation}

\indent
Noiselets totally spread out the signal energy in the measurement domain and are known to be maximally incoherent with the Haar wavelet. The mutual incoherence parameter between the noiselet measurement matrix $\Phi$ and the sparsifying Haar wavelet transform matrix $\Psi$ has been shown to be equal to 1 \cite{candes2007sparsity}, which is the minimum value possible for the incoherence. Therefore, theoretically, noiselets are the best suited measurement basis function for CS-MRI when the wavelet is used as  sparsifying transform matrix.

\subsection{Motivation}
The motivations behind using noiselets as a measurement matrix in MCS-MRI are as follows:
\begin{itemize}
\item Noiselet basis function is unitary and hence does not amplify noise as in the case of random encoding \cite{haldar2011compressed}.
\item Noiselets completely spreads out the signal energy in the measurement domain and are maximally incoherent with wavelets.
\item Unlike random basis, noiselet basis has conjugate symmetry. Thus, this property of symmetry can be exploited by using the partial Fourier like technique.
\item Noiselets are derived in the same way as wavelets, therefore it can be modelled as a multi-scale filter-bank and can be applied in $O(n \cdot log(n))$.
\end{itemize}
\indent
We proposed to use the noiselet encoding in the phase encode (PE) direction in 2D \textcolor[rgb]{0.00,0.00,0.00}{and 3D} MR imaging. Therefore, the acquired data is noiselet encoded in the PE direction and Fourier encoded in the frequency encode (FE) direction\textcolor[rgb]{0.00,0.00,0.00}{/s}.

\subsection{Pulse Sequence Design for Noiselet Encoding}
In conventional 2D MR imaging sequences, a spatially selective RF excitation pulse is used to select the slice and the linear spatial gradients are used to encode the excited slice onto the Fourier transform space. In \cite{panych1999mr,mitsouras2004non,qu2013spread,liu2012compressive}, it is demonstrated that the spatially selective RF excitation pulse can also be used to encode the imaging volume. In \cite{panych1998digital,liu2012compressive,panych1996implementation,haldar2011compressed}, the wavelet, SVD and random encoding profiles have been implemented using the spatially selective RF excitation pulses. An analysis using the linear response model described in \cite{panych1999mr} provides a theoretical framework to design spatially selective RF excitation pulses for implementation of non-Fourier encoding. Under the small flip angle ($\leq$ 30$^\circ$) regime, the RF pulse envelope can be calculated directly by taking the Fourier transform of the desired excitation profile. However this method of designing an RF excitation pulse requires excellent RF and main field homogeneity.

To excite a noiselet profile during excitation, one can design an RF pulse envelope by directly taking the Fourier transform of the noiselet basis functions. For an image of size 256$\times$256, the noiselet measurement matrix has 256 rows and 256 columns (see \eqref{eq:nmat} for the low dimensional example). The Fourier transformation of each row of the noiselet matrix will result in 256 RF excitation pulses.
\begin{figure} [!ht]
\includegraphics[scale=1.0]{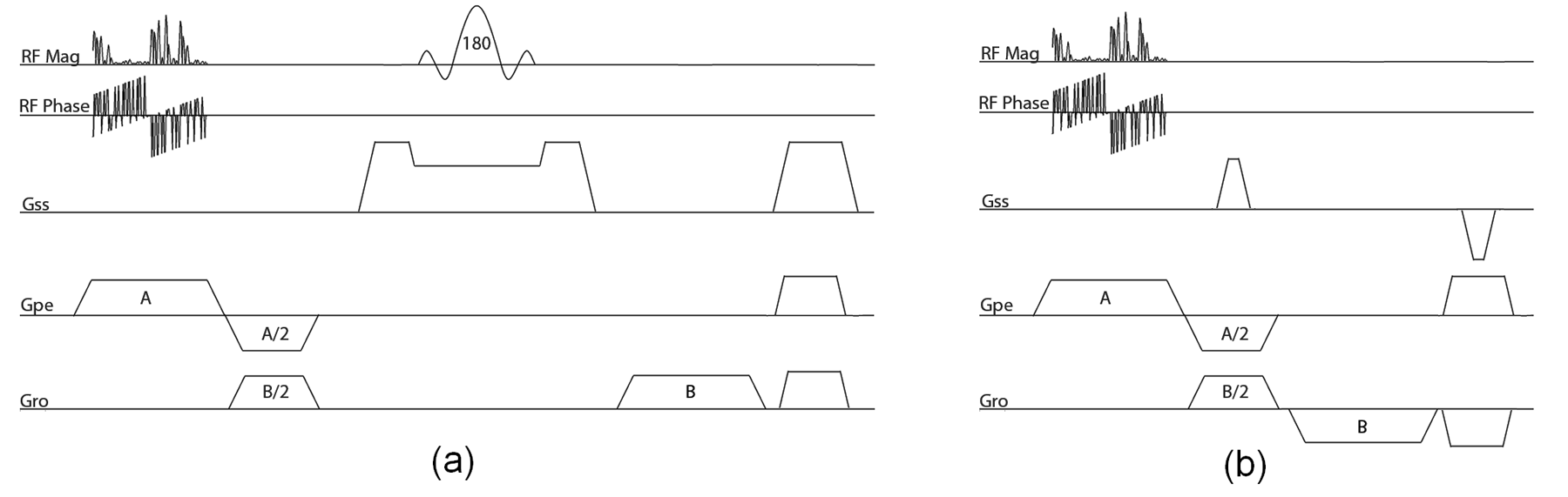}
\centering
\caption{\textbf{(a):} Pulse sequence diagram for implementation of noiselet encoding in 2D imaging, where $G_{ss}$ is the gradient in slice (z) direction, $G_{pe}$ is the gradient in phase encoding (y) direction, and $G_{ro}$ is the gradient in readout (x) direction. The RF pulse duration is 5.12 ms and the flip angle is 10$^\circ$,  which excites a noiselet profile along y-direction. The 180$^\circ$ refocusing pulse is used to select the desired slice in z-direction. A new RF pulse is executed for every new TR and the complete acquisition of all noiselet basis functions requires 256 different RF pulses derived from the noiselet measurement matrix. \textcolor[rgb]{0.00,0.00,0.00}{\textbf{(b):} Pulse sequence diagram for implementation of noiselet encoding in 3D imaging. The RF pulse excites a noiselet profile along y-direction and gradient blips are used along slice encode direction to encode slice direction with Fourier bases.}}
\label{RF}
\end{figure}

A pulse sequence for the noiselet encoding of 2D MR imaging is shown in Fig. \ref{RF} (a). The pulse sequence is designed by tailoring the spin echo sequence. The RF excitation pulse in the conventional spin echo sequence is replaced by the noiselet RF pulse, and the slice select gradient is shifted to phase encoding axis. The 180$^\circ$ refocusing RF pulse is used in conjunction with the slice selection gradient to select the slice that refocuses the spins only in the desired slice. Spoilers are used after the readout gradient to remove any residual signal in the transverse plane. A new RF excitation pulse is used for every new TR to excite a new noiselet profile, and a total of 256 TR are required for excitation of the complete set of noiselet bases. The readout gradient strength determines the FOV in the readout direction, while the phase encoding gradient strength and duration of the RF excitation pulse determines the field of view (FOV) in phase encoding direction. The FOV in phase encoding direction is determined as
\begin{equation}
\label{eq:FOV_pe}
FOV_{pe} = \frac{1}{\gamma G_y \Delta t_{p} }
\end{equation}
where $G_y$ is the gradient strength in PE direction, $\gamma$ is the gyromagnetic ratio, and $\Delta t_{p}$ is the dwell time of the RF pulse which is defined as $\Delta t_{p}$ = (Duration of RF pulse) / (Number of points in RF pulse). Equation \eqref{eq:FOV_pe} is used to calculate the gradient strength $G_y$ required in the phase encoding direction during execution of RF excitation pulse.

\textcolor[rgb]{0.00,0.00,0.00}{The method described above can also be used to design the pulse sequence for the noiselet encoding of 3D MR imaging as shown in Fig. \ref{RF} (b).}

\subsection{Under-sampling in noiselet encoding}

\begin{figure} [!ht]
\includegraphics[scale=1.0]{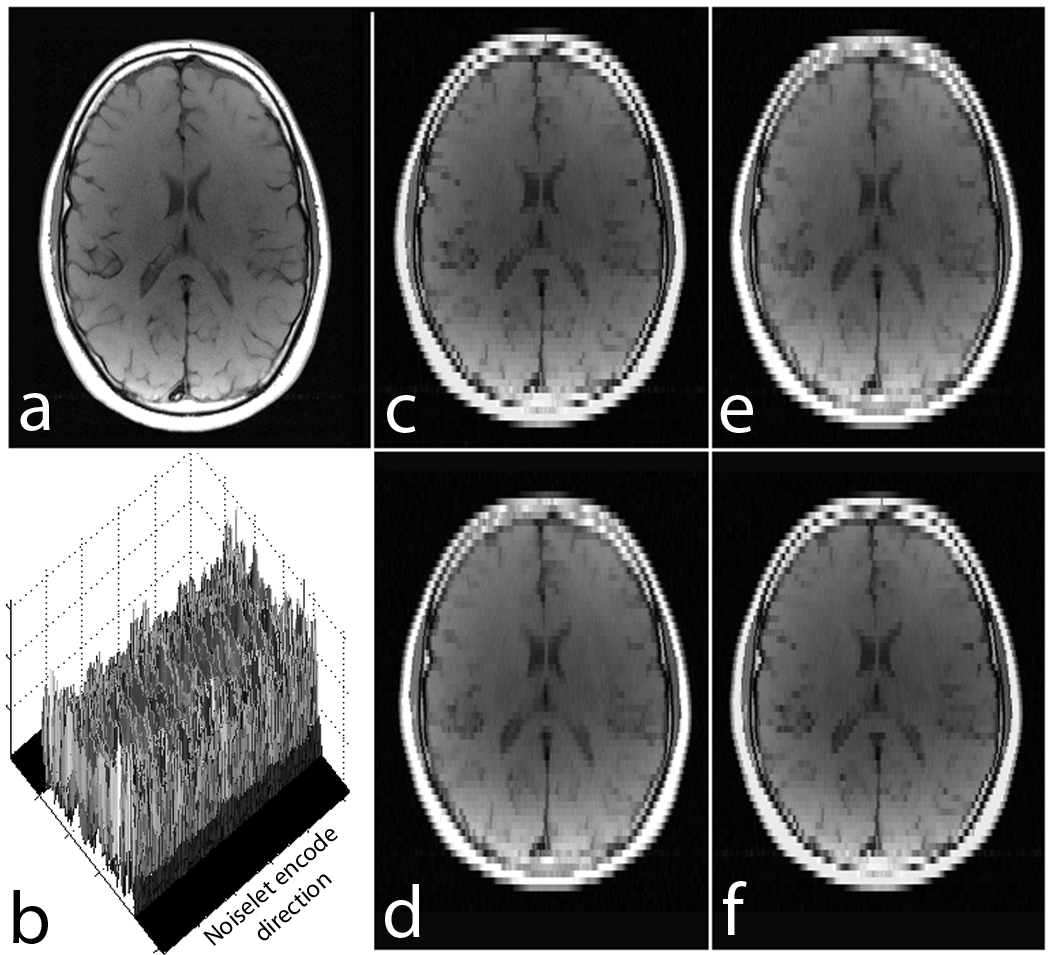}
\centering
\caption{\textbf{(a):} Brain image of size $256\times256$; \textbf{(b):} 3D magnitude map of the noiselet transform of the brain image along phase encoding direction (noiselet encodes); \textbf{(c):} image reconstructed using only the first 64 noiselet encodes; \textbf{(d), (e) and (f):} are the images reconstructed with the second, third and fourth 64 noiselet encodes respectively.}
\label{ss}
\end{figure}

Noiselet transform is a type of Haar-Walsh transform. The noiselet transform coefficients totally spread out the signal in scale and time (or spatial location) \cite{coifman2001noiselets}. As a result, each subset of the transform coefficients contains a certain information of the original signal at all the scales and times (spatial locations), and can be used alone with zero padding to reconstruct the original signal at a lower resolution. This important property is demonstrated by the example shown in Fig. \ref{ss}.

Fig. \ref{ss} shows a brain image of size $256\times256$, and the 3D magnitude map of the noiselet transform of the brain image along the phase encoding direction (all noiselet encodes). Fig. \ref{ss} (c-f) shows the images reconstructed with the first, second, third and fourth 64 noiselet encodes by zero padding the rest. Each of these images are reconstructed using one quarter of the noiselet encodes and has low resolution than the original image. However, each of these images have complementary information about the original image and have approximately the same amount of energy and information because they are reconstructed using the same size of partial matrix from the original coefficient matrix.

\begin{figure} [!ht]
\includegraphics[scale=1.0]{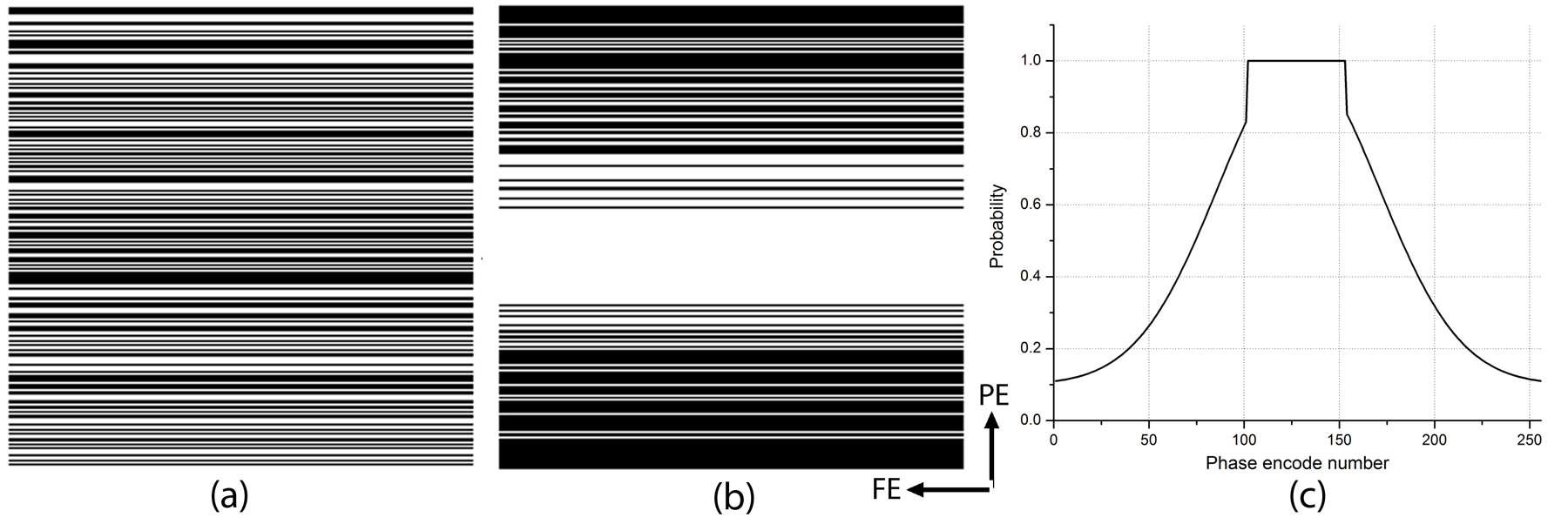}
\centering
\caption{\textbf{(a):} Completely random sampling pattern for noiselet encoding generated using the uniform probability distribution function, where the white lines represent the sampled phase encodes; \textbf{(b):} variable density random under-sampling pattern for the Fourier encoding scheme, with the sampling mask generated according to Gaussian probability distribution function; and
\textcolor{black}
{ % color begins
\textbf{(c):} probability distribution function used to generate variable density random undersampling pattern where the probability of sampling the center phase encodes is equal to 1, while the probability decays as a Guassian function with distance from the center phase encodes. The central fully sampled region is always between 20\%-25\% of the total number of sampled phase encodes.}
} % color ends
\label{mask}
\end{figure}

Based on the above property of noiselet transform, we propose to under-sample the noiselet encoded data along the phase encoding direction according to the uniform probability distribution function. One sampling mask using this scheme is shown in Fig. \ref{mask}(a) where the white lines represent the sampled data points and the black lines represent the unsampled data points. \textcolor[rgb]{0.00,0.00,0.00}{Fig. \ref{mask}(b) shows the sampling mask for Fourier encoding scheme drawn from a variable density probability distribution function shown in Fig. \ref{mask}(c).
}
\textcolor{black}
{ % color begins
\subsection{Empirical RIP analysis of measurement matrix}
According to the CS theory summarized in Section II, the measurement matrix is crucial to the performance of CS reconstruction, and the performance of a measurement matrix $\Phi$ for a $K$-sparse signal $x$ can be assessed by the statistics of $\sigma_{min}[\Phi_{sub}(K)]$ and $\sigma_{max}[\Phi_{sub}(K)]$ over the $\Phi_{sub}(K)$s consisting of $k$ distinct columns of $\Phi$. To understand the behavior and advantage of the noiselet encoding proposed above,  we have used this method to assess the noiselet measurement matrix in comparison to the conventional Fourier measurement matrix.
}

\textcolor{black}
{ % color begins
In the assessment, the size of the signal was $n = 256$, the number of measurements $m =100$, the number of channels $L = 1, 8$ and $14$, and the sparsity $K$ was varied from 5 to 100 with an increment of 5. For $L = 1$, the measurement matrices, $\Phi$s, were generated for the noiselet and Fourier encodings, respectively. For $L = 8$ and $14$, the measurement matrices $E$s as given in (\ref{eq:acq_pmri}) were generated for the noiselet and Fourier encodings, respectively. For each $K$, 2,000 submatrices $\Phi_{sub}(K)$s  were drawn uniformly randomly from the columns of $\Phi$, then the $\sigma_{min}[\Phi_{sub}(K)]$ and $\sigma_{max}[\Phi_{sub}(K)]$ of every $\Phi_{sub}(K)$ were calculated. The same procedure is used to obtain the submatrices $E_{sub}(K)$s from $E$ and to calculate the $\sigma_{min}[E_{sub}(K)]$ and $\sigma_{max}[E_{sub}(K)]$ of every $E_{sub}(K)$. The statistics of  $\sigma_{min}[\Phi_{sub}(K)]$s, $\sigma_{max}[\Phi_{sub}(K)]$s, $\sigma_{min}[E_{sub}(K)]$s and $\sigma_{max}[E_{sub}(K)]$s were accumulated from their respective 2,000 samples.
}

\begin{figure} [!ht]
\includegraphics[scale=1.0]{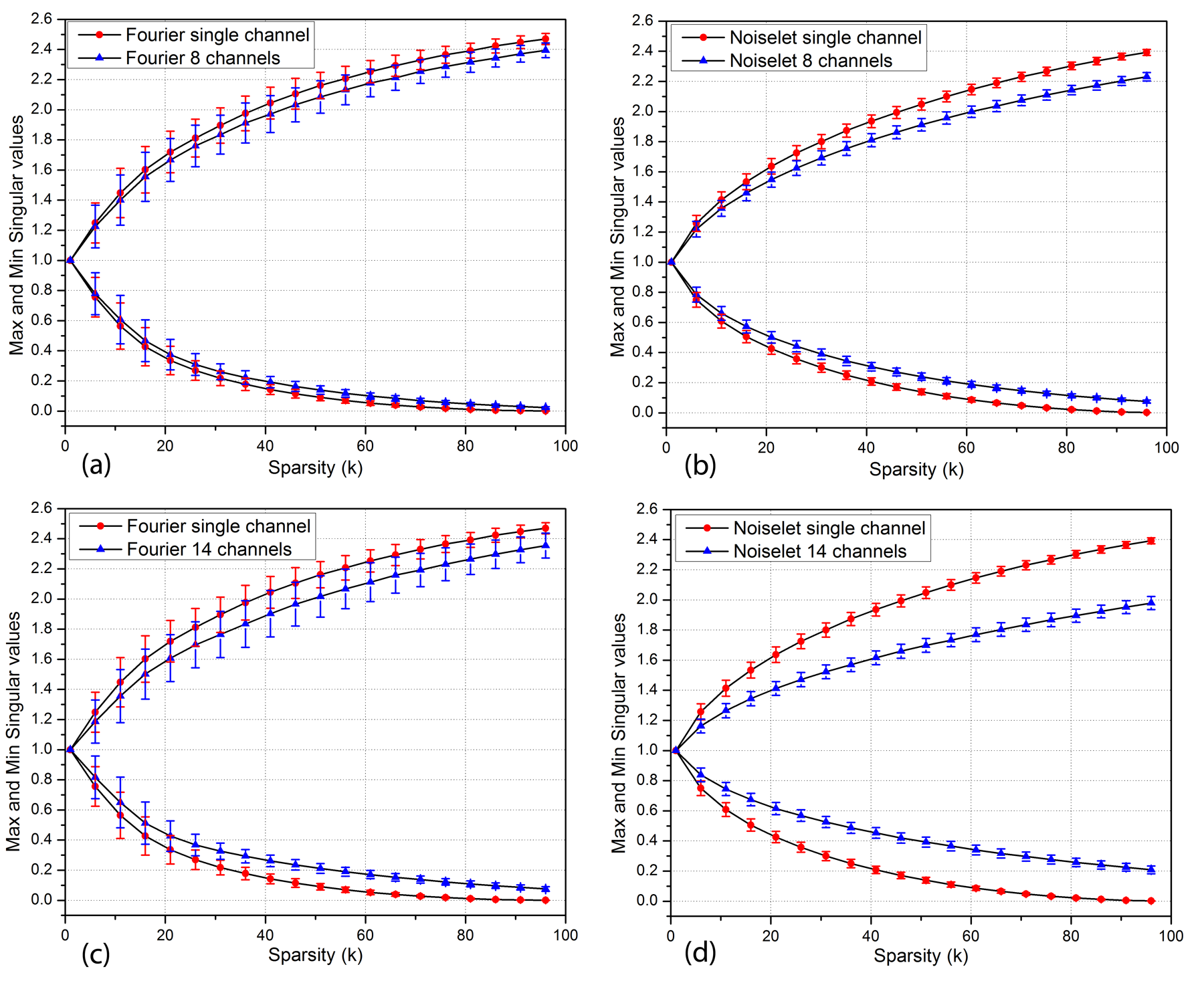}
\centering
\caption{\textcolor{black}{The means and standard deviations of maximum and minimum singular values versus sparsity $K$ for {\textbf{(a)}} and  {\textbf{(c):} Fourier measurement matrix,} {\textbf{(b)}} and {\textbf{(d):}} noiselet measurement matrix.} 
}
\label{CN}
\end{figure}

\textcolor{black}
{% color begins
Fig. \ref{CN} shows the means and standard deviations of the minimum and maximum singular values of $\Phi_{sub}(K)$s and $E_{sub}(K)$s versus the sparsity $K$ for the Fourier and noiselet measurement matrices. As seen from the figure, in single channel case, the singular values of noiselet measurement matrix are closer to 1 than those of Fourier measurement matrix, but are not significantly different. As the number of channels increases, the singular values of noiselet and Fourier measurement matrices all move towards 1, but those of noiselet measurement matrix move much closer to 1 than those of Fourier measurement matrix. By the CS theory, when the maximum distance from 1 to the singular values is less than 1, it equals roughly the RIP constant $\delta_K$. Therefore, the figure actually reveals two facts: 1) For both the noiselet and Fourier measurement matrices, the RIP constant $\delta_K$ decreases as the number of channels increases. 2) As the number of channels increases, the RIP constant $\delta_K$ of noiselet measurement matrix deceases much more than that of Fourier measurement matrix. According to the CS theory, these imply that the multichannel measurement matrix should generally outperform the single channel measurement matrix, and that the multichannel noiselet measurement matrix should generally outperform the multichannel Fourier measurement matrix.
}

\textcolor{black}
{% color begins
As a particular example consider the curves in Fig. \ref{CN} (d) for the noiselet measurement matrix. To facilitate discussion, the distances from 1 to the singular values of a measurement matrix will be called the $\delta$-distances here. In single channel case, the $\delta$-distances of noiselet measurement matrix are less than 1 for $K \leq 40$. By RIP, this implies that the single channel noiselet measurement matrix can guarantee the recovery of the signals with sparsity $K \leq 20$. When the number of channels is increased to 14, the $\delta$-distances are less than 1 for $K \leq 85$. So the 14 channel noiselet measurement matrix can guarantee the recovery of the signals with sparsity $K \leq 42$. The improvement in terms of sparsity is two folds. In contrast, for the 14 channel Fourier measurement matrix shown in Fig. \ref{CN} (c), its $\delta$-distances are less than 1 only for $K < 30$, so it can only guarantee the recovery of the signals with sparsity $K < 15$.
} % color ends

\textcolor{black}{From the above assessment, it can be expected that the multichannel CS MRI will outperform the single channel CS MRI and that the noiselet encoding multichannel CS MRI will outperform the Fourier encoding multichannel CS MRI in practice. These are confirmed by the simulation and experiment results presented in the next sections.}

\textcolor{black}{
It is important to note that the above empirical analysis results are obtained by using the complex valued sensitivity maps. If only the magnitudes of the sensitivity maps are used, the above results will not hold. Therefore, the complex valued sensitivity maps of multiple receive coils are the source for the significant performance improvement of the multichannel measurement matrix.
}

%\begin{figure} [!ht]
%\includegraphics[scale=0.1]{condition_number.png}
%\centering
%\caption{\textcolor{black}{ \textbf{(a)} Plot of mean condition number w.r.t sparsity (k) for Fourier measurement matrix; \textbf{(a)} Plot of mean condition number w.r.t sparsity (k) for noiselet measurement matrix. The improvement in contion number for Fourier encoding is 25\% when going from single channel to mutiple channel while for noiselet measurement matrix the improvement is 50\%. Therefore it can be inffered that noiselet encoded MCS would perform better than the Fourier encoded MCS.}}
%\label{CN}
%\end{figure}

\section{Simulation Study and Results}
Simulations were performed on a (256$\times$256) brain image  to investigate the performance of noiselet encoded and Fourier encoded MRI. The simulation study was divided into two parts: (i) a simulation study with a single channel using uniform sensitivity and (ii) a simulation study with multiple channels where the sensitivity profiles were estimated from the acquired data.

\subsection{Single Channel Simulation with a Uniform Sensitivity Profile}
\noindent
\textbf{Fourier encoded CS-MRI} A Fourier transform of the image was taken in the PE direction to simulate Fourier encoding. Two types of sampling strategies were used to sample Fourier encoded data: (i) a variable density random sampling pattern as shown in Fig. \ref{mask} (a) where samples were taken in the PE direction according to a Gaussian distribution function, and (ii) a completely random sampling pattern as shown in Fig. \ref{mask} (b) where samples were taken in the PE direction according to the uniform density function. The non-linear program of \eqref{eq:min_tv_wav} was solved to reconstruct the final image for acceleration factors of 2 and 3. In these cases the encoding matrix $E$ does not have any sensitivity information (i.e. $E = \Phi$).

\noindent
\textbf{Noiselet encoded CS-MRI} A noiselet transform of the image was taken in the PE direction to simulate noiselet encoding. A completely random sampling pattern was used to sample the noiselet encoded data in the PE direction and the non-linear program of \eqref{eq:min_tv_wav} was solved to reconstruct the final image for acceleration factors of 2 and 3. In these cases the encoding matrix $E$ does not have any sensitivity information (i.e. $E = \Phi$).

\begin{figure} [!ht]
\includegraphics[scale=1.0]{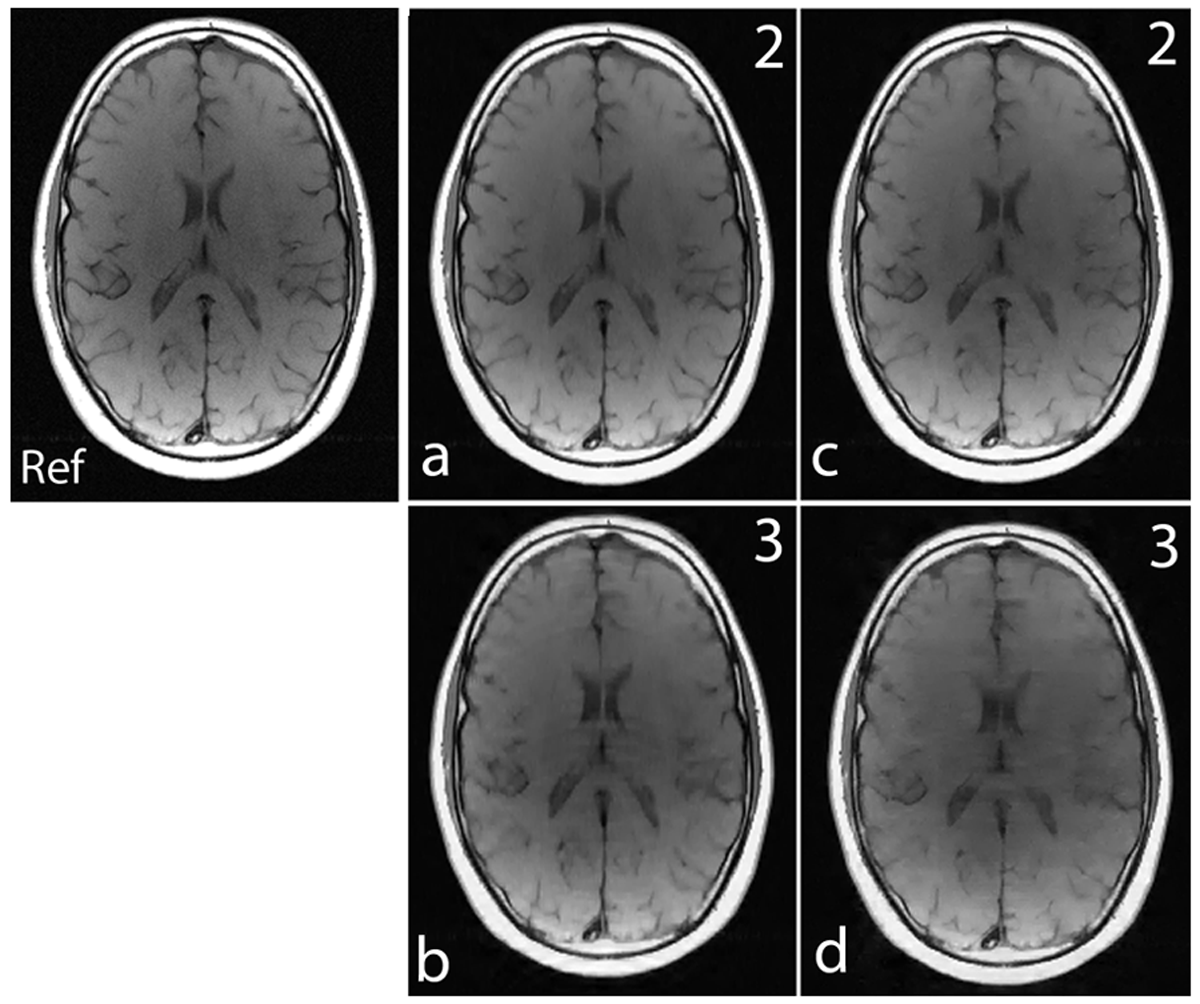}
\centering
\caption{\textbf{Ref:} represents the reference image 256$\times$256 (up/down: phase encodes, left/right: frequency encode); \textbf{(a)-(b):} show images reconstructed using Fourier encoding with variable density random under-sampling patterns for acceleration factors of 2 and 3 respectively; \textbf{(c)-(d):} show images reconstructed using noiselet encoding with completely random under-sampling patterns for acceleration factors of 2 and 3 respectively. Noiselet encoded CS-MRI performs better than the Fourier encoded CS-MRI when completely random under-sampling is used for both the encoding schemes. This is due to the better incoherence provided by the noiselets. However, noiselet encoding with a random under-sampling pattern performs similar to Fourier encoding with a variable density random under-sampling pattern.}
\label{NF_single}
\end{figure}

\indent
Fig. \ref{NF_single} show the images reconstructed with Fourier encoding and noiselet encoding using variable density random under-sampling and completely random under-sampling pattern respectively. The noiselet encoded CS-MRI performs similar to that of the Fourier encoded CS-MRI. This is due to the fact that in the case of variable density random under-sampling, the Fourier encoding judiciously exploits extra information about the data, namely the structure of k-space. The center of the k-space data has maximum energy and hence, by densely sampling the center of k-space, the Fourier encoding captures most of the signal energy and results in better performance.

In practice, the MR data is collected through the use of multiple channels, and data in each channel is slightly different from the other channels. The actual k-space data is convolved with the Fourier transform of the sensitivity profiles of the individual channel, making the data from each channel different from others. This sensitivity information can also be taken into consideration while performing the CS reconstruction, by applying the multichannel CS frame. Therefore,  to further study the effect of sensitivity information on noiselet encoding and Fourier encoding, MCS-MRI simulations were performed. To quantitatively compare the performance of both the encoding schemes, we used the relative error defined in \eqref{eq:rel_err} as a metric:
\begin{equation}
\label{eq:rel_err}
Relative\hspace{1mm}error = \frac{\|x_0-\hat{x}\|_{l_2}}{\|x_0\|_{l_2}}
\end{equation}

\begin{figure} [!ht]
\includegraphics[scale=1.0]{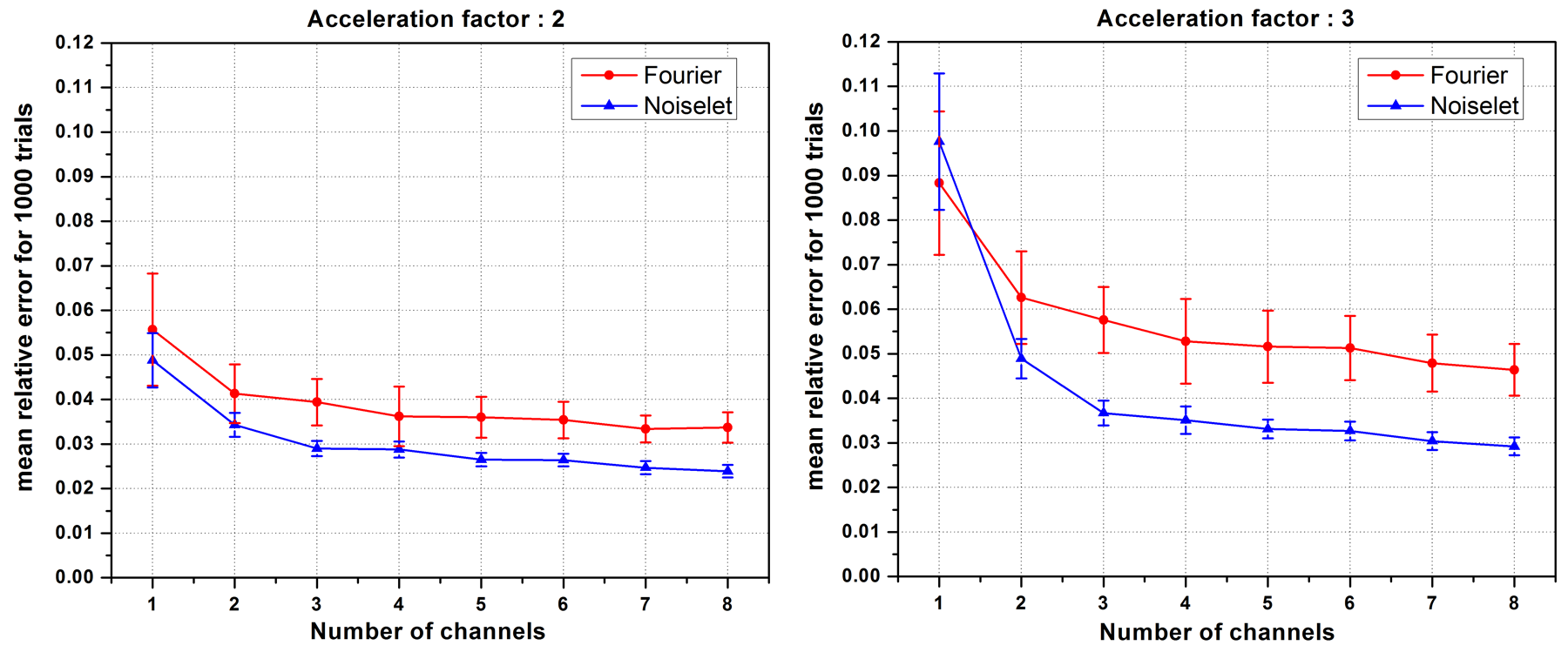}
\centering
\caption{\textcolor[rgb]{0.00,0.00,0.00}{The mean relative error and standard deviation} (vertical bar) versus the number of receive channels for acceleration factors of 2 and 3, showing that the error increases as the number of channels decreases. Noiselet encoding outperforms Fourier encoding for both acceleration factors when the number of channels is more than two. However for a single channel, noiselet encoding outperforms Fourier encoding only for the acceleration factor of 2.}
\label{err_nch}
\end{figure}

\indent
First we investigated the effect of the number of channels on the reconstruction quality using the MCS framework. For a fixed number of measurements, the number of channels was varied and the mean of the relative error for 1000 such simulations was calculated. Fig. \ref {err_nch} shows the plot of the mean relative error versus the number of channels for the acceleration factors of 2 and 3. When the number of channels was two, the noiselet encoding scheme outperformed the Fourier encoding scheme for both the acceleration factors of 2 and 3. However, when number of channels was equal to one, the noiselet encoding outperformed the Fourier encoding for an acceleration factor of 2, but not for an acceleration factor of 3. It is interesting to note that noiselet encoding outperformed Fourier encoding for both acceleration factors when the number of channels was greater than one. These simulations suggest that noiselet encoding should take into account the sensitivity information while performing CS, and therefore noiselet encoding is potentially a better encoding scheme for MCS-MRI. Based on the fact that noiselet encoding performs better than Fourier encoding for multi-channel data, we investigated the performance of both the encoding schemes using muti-channel data.

\subsection{Multiple Channel Simulation}
\begin{figure} [!ht]
\includegraphics[scale=1.0]{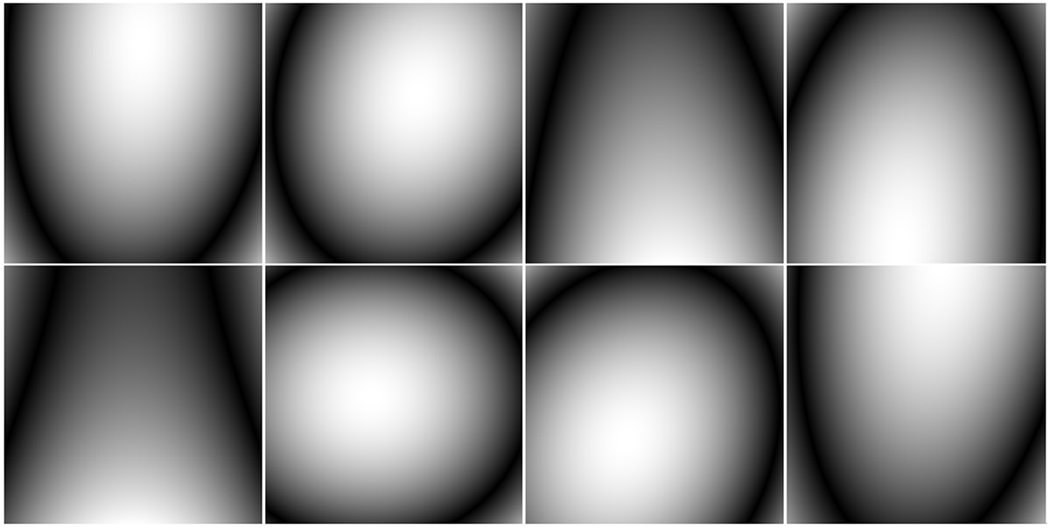}
\centering
\caption{The eight coil sensitivity magnitude maps used in simulations that were estimated from the data acquired on MR scanner.}
\label{maps}
\end{figure}

A (256$\times$256) brain image  was used to compare the performance of noiselet encoding and Fourier encoding in MCS-MRI for different acceleration factors. Eight complex sensitivity maps (Fig. \ref{maps}) obtained from the head coil of a Siemens Skyra 3T scanner were used to perform the simulations. For solving the minimization program in \eqref{eq:min_tv_wav}, we used the nonlinear conjugate gradient with the backtracking line search method \cite{lustig2007sparse}. The measurement matrix ($\Phi$) was the discrete Fourier transform matrix while the daubechies-4 wavelet transform matrix ($\Psi$) and TV were used as sparsifying transforms.

\textbf{Fourier encoded MCS-MRI} The reference brain image was multiplied by the sensitivity function to generate eight sensitivity encoded images. The Fourier transform of each these images was taken in the PE direction; only a few PEs were taken according to the Gaussian probability distribution function. MCS-MRI reconstruction of \eqref{eq:min_tv_wav} was solved using the nonlinear conjugate gradient on this data. An example sampling scheme for the Fourier encoded MCS-MRI is shown in Fig. \ref{mask}(a).
\textbf{Noiselet encoded MCS-MRI} A Noiselet transform of the sensitivity encoded images was taken in the PE direction, with only a few PE selected according to the uniform probability distribution function. MCS-MRI reconstruction of \eqref{eq:min_tv_wav} was solved using the nonlinear conjugate gradient on this data. An example of the sampling scheme for noiselet encoded MCS-MRI is shown in Fig. \ref{mask}(b).

\begin{figure}
\centering
\includegraphics[scale=0.3]{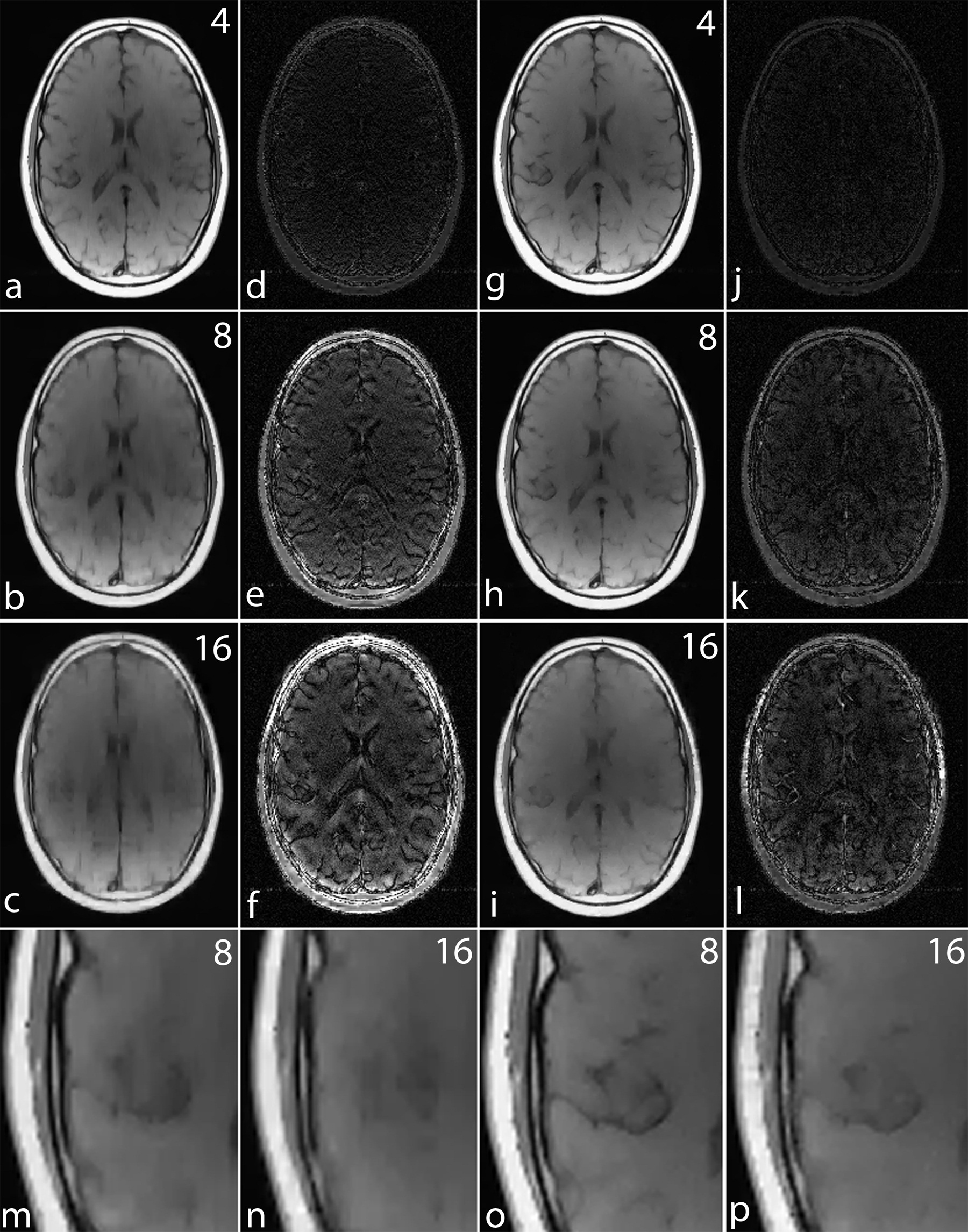}
\centering
\caption{Simulation results for MCS-MRI comparing the noiselet encoding and Fourier encoding schemes (up/down: phase encodes, left/right: frequency encode). \textbf{(a)-(c): } show images reconstructed with Fourier encoding for acceleration factors of 4, 8, and 16 respectively; \textbf{(d)-(f): } show difference images with Fourier encoding for acceleration factors of 4, 8, and 16 respectively; \textbf{(g)-(i): } show images reconstructed with noiselet encoding for acceleration factors of 4, 8, and 16 respectively; \textbf{(j)-(l): } show difference images with noiselet encoding for acceleration factors of 4, 8, and 16 respectively; \textbf{(m)-(n): } show zoomed portion of images reconstructed with Fourier encoding for acceleration factors of 8, and 16 respectively; \textbf{(o)-(p): } show zoomed portion of images reconstructed with noiselet encoding for acceleration factors of 8, and 16 respectively. The zoomed images highlight that MCS-MRI using noiselet encoding reconstructions outperforms the Fourier encoding for preserving image resolution.}
\label{brain_recon}
\end{figure}

\indent
For a noiseless simulation, the reconstructed images for different acceleration factors (4, 8 and 16) are shown in Fig. \ref{brain_recon}. The difference images in Fig. \ref{brain_recon} (d)-(f) and (j)-(l) demonstrate that  the error in noiselet encoding is always less than in Fourier encoding, and that the noiselet encoded MCS-MRI reconstruction preserves spatial resolution better than the Fourier encoded MCS-MRI. Fig. \ref{brain_recon} (m) and (n) show the zoomed images reconstructed with Fourier encoding for acceleration factors of 8 and 16 respectively, while Fig. \ref{brain_recon} (o) and (p) show the zoomed images reconstructed with noiselet encoding for an acceleration factors of 8 and 16 respectively. The zoomed images highlight that the spatial resolution of the noiselet encoded reconstructions outperforms the Fourier encoded reconstructions. Moreover, the spatial resolution provided by the noiselet encoding at an acceleration factor of 16 is comparable to that of the Fourier encoding at an acceleration factor of 8, suggesting that noiselet encoding performs approximately twice as good as Fourier encoding.

\begin{figure} [!ht]
\includegraphics[scale=1.0]{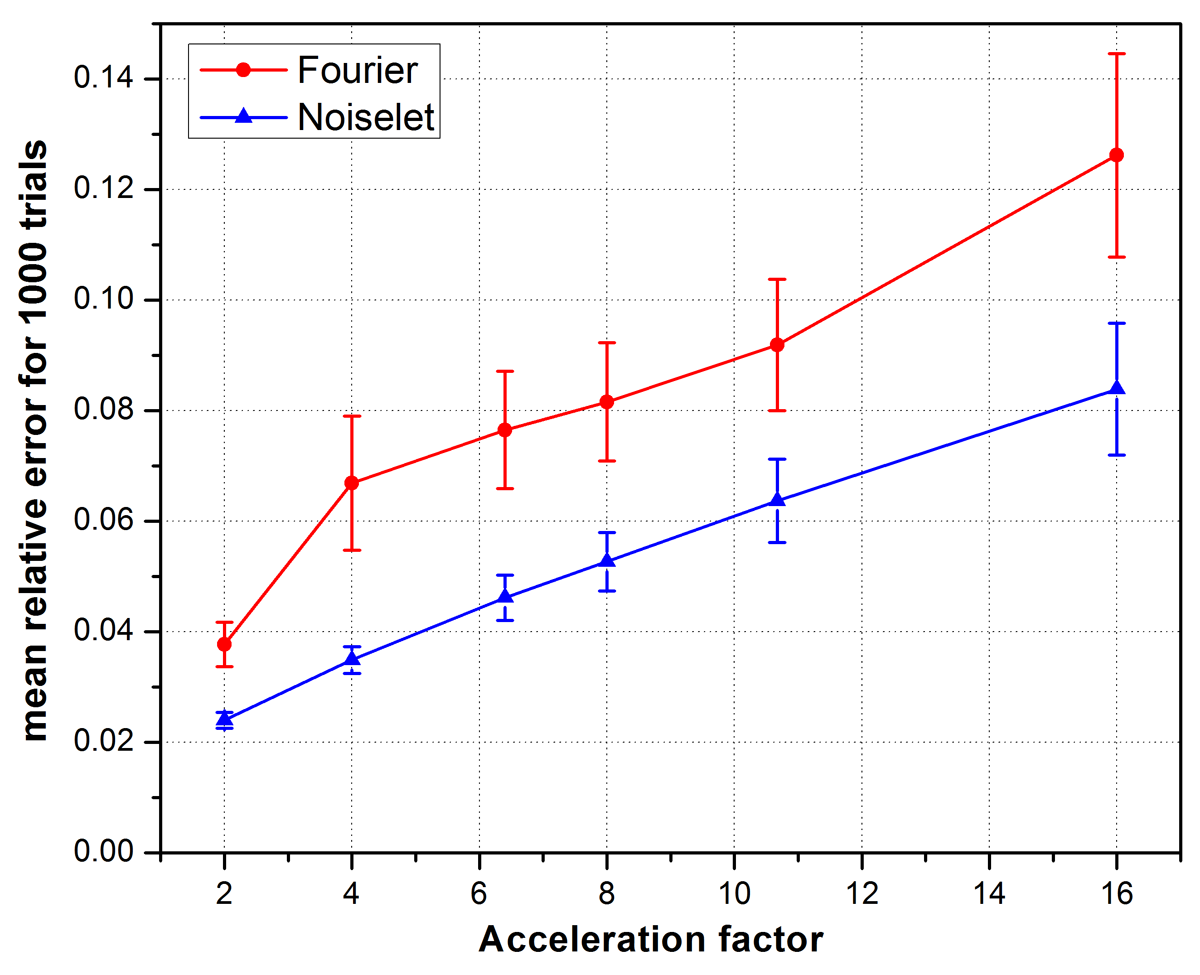}
\centering
\caption{\textcolor[rgb]{0.00,0.00,0.00}{The mean relative error and standard deviation }(vertical bar) versus the acceleration factor in MCS-MRI highlighting that noiselet encoding consistently outperforms Fourier encoding.}
\label{err_meas}
\end{figure}

To measure the relative error, simulations were performed on the brain image for 1000 times by randomly generating a sampling mask each time. The mean of the relative errors was calculated after 1000 such reconstructions at every acceleration factor. The mean relative error versus the number of measurements is plotted in Fig. \ref {err_meas} and highlights that noiselet encoding outperforms Fourier encoding for all acceleration factors. The relative error for noiselet encoding at an acceleration factor of 16 was the same as the relative error for Fourier encoding at an acceleration factor of 8 indicating that higher acceleration factors are achievable with noiselet encoding compared to Fourier encoding.

\begin{figure}
\centering
\includegraphics[scale=1.0]{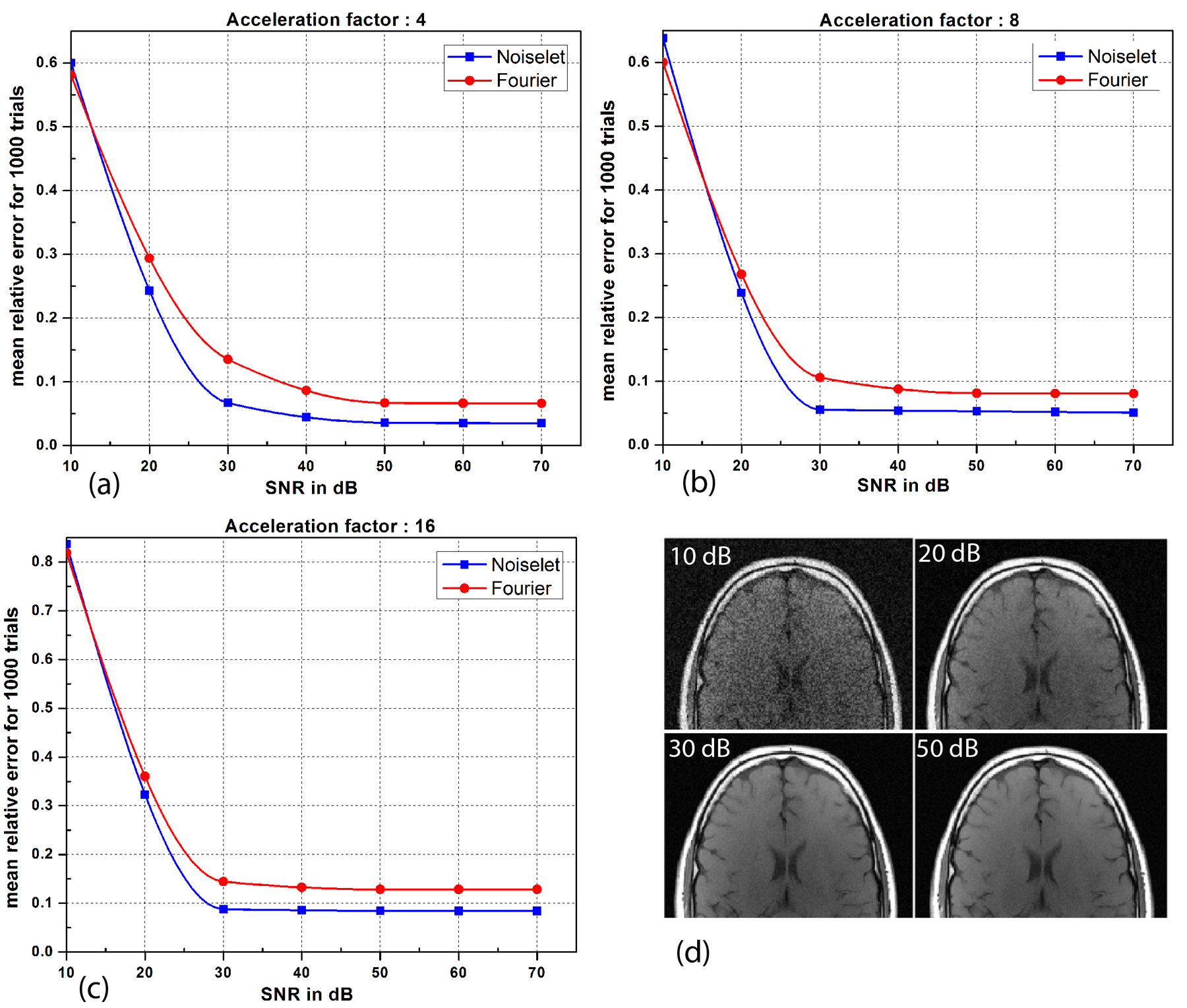}
\centering
\caption{\textbf{(a), (b) and (c):} are the plots of the mean relative error as a function of the signal to noise ratio (SNR) for different number of measurements. When the SNR is greater than 20 dB, the noiselet encoding outperforms Fourier encoding in the presence of noise for all acceleration factors; \textcolor[rgb]{0.00,0.00,0.00}{\textbf{(d):} show the brain images with SNR of 10, 20, 30 and 50 dB.}}
\label{err_snr}
\end{figure}

\indent
In practice, MR data always has some noise and the level of noise depends upon many factors including the FOV, resolution, type of imaging sequence, magnetic field inhomogeneity and RF inhomogeneity. Therefore, simulations were carried out to evaluate the performance of both the noiselet encoding and Fourier encoding schemes in the presence of variable levels of noise. Different levels of random Gaussian noise were added to the measured k-space data, and MCS-MRI reconstruction was performed for noiselet and Fourier encoding schemes. For every level of noise, 1000 simulations were performed and the mean of the relative error was calculated. Fig. \ref{err_snr} shows the mean relative error as a function of the Signal to Noise Ratio (SNR), demonstrating the comparative performance of noiselet encoding reconstructions and Fourier encoding reconstructions in the presence of noise. The plots demonstrate that noiselet encoding outperforms Fourier encoding for SNR above 20 dB for all acceleration factors, but does a poor job at SNR of 10 dB. \textcolor{black}{However, as shown in the images in Fig. \ref{err_snr}(d), only the images with SNR above 20 dB are adequate for diagnostic purposes. Thus, acquisitions with  10 dB SNR is not a viable scanner operation condition and the poor performance of noiselet encoding at 10 dB SNR is not a practical limitation. The poor performance of noiselet encoding at 10 dB SNR can be attributed to the fact that at extremely low SNR, most of the noiselet coefficients are severely corrupted by the noise since their magnitudes are approximately uniform. In contrast, the Fourier coefficients at the center of k-space have much larger magnitudes and hence are less affected by the noise at low SNR. These large magnitude coefficients are fully utilized in reconstruction because of the centralized variable density sampling scheme, hence Fourier encoding is less affected by the noise and performs better at low SNR.}

\section{Experiments}
Experiments were carried out on Siemens Skyra 3T MRI scanner with a maximum gradient strength of  40 mT/m and a maximum slew rate of 200 mT/m/sec. Informed consent was taken from healthy volunteers in accordance with the Institution's ethics policy. To validate the practical implementation of noiselet encoding, the pulse sequence shown in Fig. \ref{RF} was used to acquire noiselet encoded data. An RF excitation pulse with 256 points was used with the duration equal to 5.12 ms and  flip angle of 10$^\circ$. We also acquired the Fourier encoded data  using the spin echo (SE) sequence to compare the quality of the reconstructed image from the data acquired by the noiselet encoding sequence. An apodized slice selective sinc RF excitation pulse was used in the spin echo sequence with a duration of 2.56 ms and a flip angle of 10$^\circ$. The  protocol parameters for the noiselet encoding sequence and the Fourier encoding SE sequence were \textbf{(i) Phantom experiments} FOV = 200 mm, TE/TR = 26/750 ms, averages = 2, image matrix = 256$\times$256; and \textbf{(ii) \textit{In vivo} experiments} FOV = 240 mm, TE/TR = 26/750 ms, averages = 2, image matrix = 256$\times$256.

\textcolor{black}
{
%\subsection{Sensitivity map estimation}
The performance of the MCS-MRI reconstruction depends on the accuracy of the sensitivity matrix estimated. We used the regularized self-calibrated estimation method \cite{kim2008smoothing} to estimate the sensitivity maps from the acquired data. This method estimates the sensitivity map $\hat{\Gamma}_i$ of the $i$th receive coil by using
\begin{equation}
\label{eq:sense_est}
\hat{\Gamma}_i = \Min_{\Gamma} \frac{1}{2} \|I_i-\Gamma I_{ref}\|^2 + \beta R(\Gamma)
\end{equation}
where $i \in [1,2, \cdots, L]$ and $ R(\Gamma)$ is a spatial roughness penalty function with weighting factor $\beta$. The reference image $I_{ref}$ can be obtained using the sum of squares of individual coil images $I_i$'s. For the experimental results presented below, the sensitivity maps were estimated from fully sampled images using \eqref{eq:sense_est}. The data was acquired using a 32 channel head coil, but only the data from 14 channels with good SNR was used in the reconstruction.
}
\begin{figure} [!ht]
\includegraphics[scale=0.3]{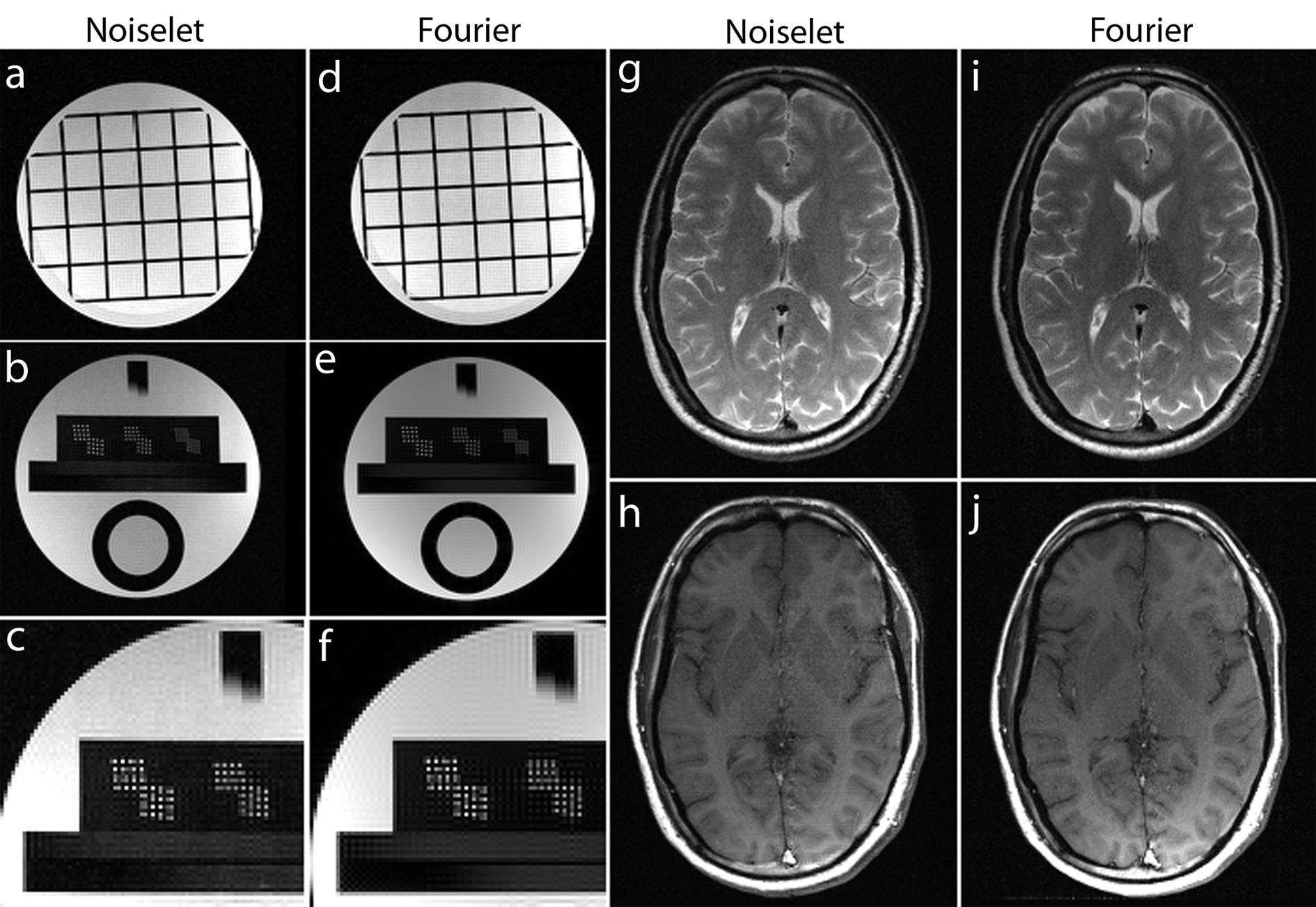}
\centering
\caption{Images reconstructed using fully sampled noiselet encoded and Fourier encoded data acquired on the 3T scanner (up/down: phase encodes, left/right: frequency encode). The noiselet encoded data was acquired using the pulse sequence described in section III C, and Fourier encoded data was acquired using a conventional spin echo sequence. \textbf{(c)-(f):} show the zoomed portion of the images in (b) and (e) respectively, with the zoomed images demonstrating that noiselet encoding provides similar image resolution to that of Fourier encoding; \textbf{(g)-(h):} show T2 and T1 weighted brain images using noiselet encoding respectively; \textbf{(i)-(j):} show T2 and T1 weighted brain images using Fourier encoding respectively. These \textit{in vivo} images demonstrate the practical feasibility of the proposed noiselet encoding scheme.}
\label{F_N}
\end{figure}

Non-Fourier encoding in general is sensitive to field inhomogeneities, but careful design of the sequence and good shimming  can result in high quality images. To reconstruct the noiselet encoded data the inverse Fourier transform was taken along the frequency encoding axis and the inverse noiselet transform was taken along the PE axis. To reconstruct the Fourier encoded data, an inverse Fourier transform was taken along both axes. Fig. \ref{F_N} shows the images reconstructed from the noiselet encoded data and Fourier encoded data sets. These images demonstrate that the noiselet encoding reconstructions are practically feasible and produce artifact free images. Fig. \ref{F_N}(c) shows a zoomed portion of the noiselet encoded image, while Fig. \ref{F_N}(f) shows a zoomed portion of the Fourier encoded image. The zoomed images reveal that the resolution of the image from noiselet encoding with 256 noiselet excitation is the same as that of the image from Fourier encoding with 256 phase encodes. Fig. \ref{F_N} (g) and (i) show the T2 weighted images for the brain with noiselet encoding and Fourier encoding, respectively. Fig. \ref{F_N} (h) and (j) show the T1 weighted images for the brain with noiselet encoding and Fourier encoding, respectively. It is evident from the \textit{in vivo} images that the proposed noiselet encoding is feasible in practice.

\begin{figure} [!ht]
\includegraphics[scale=0.33]{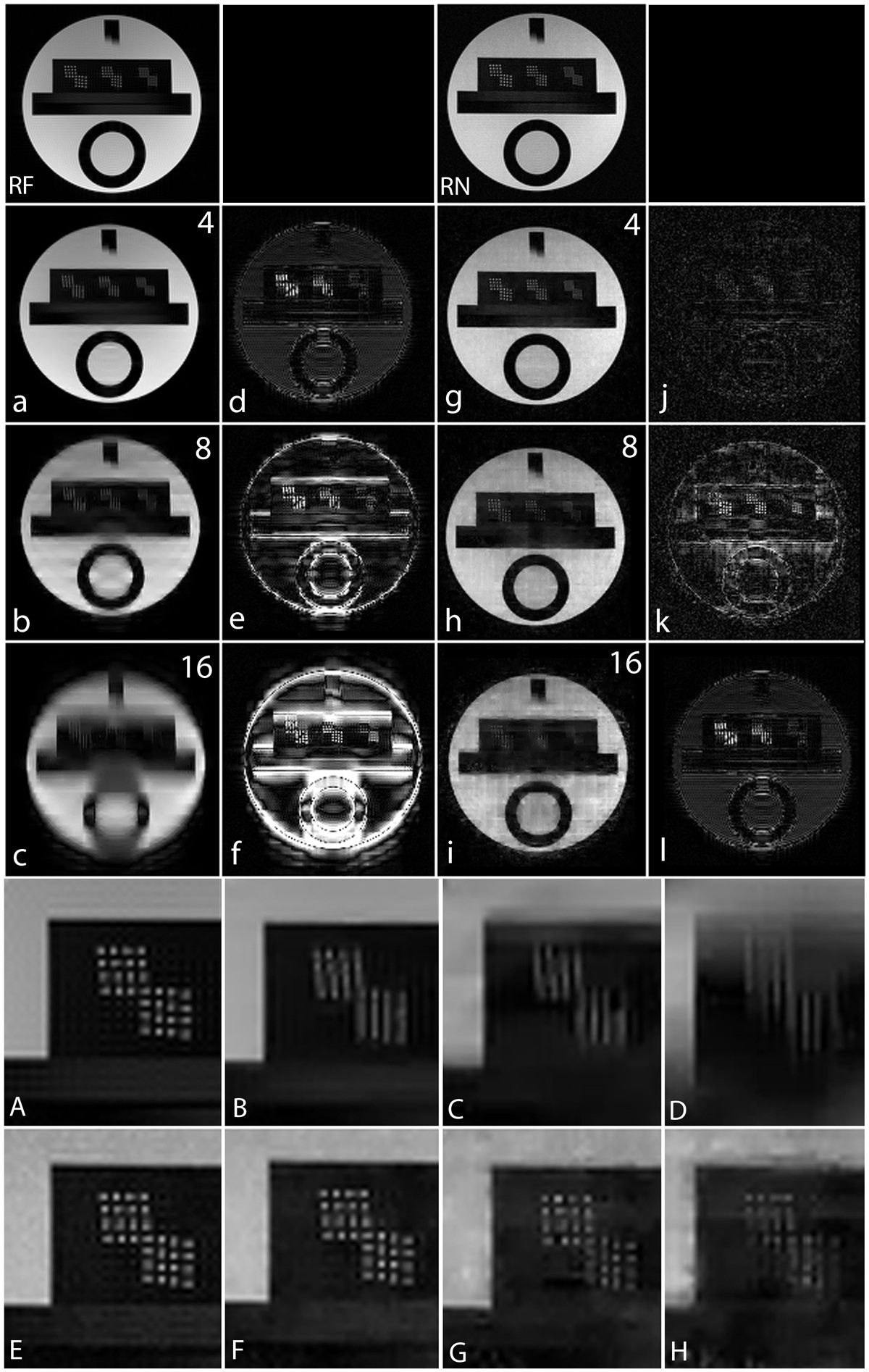}
\centering
\caption{MCS-MRI reconstruction on the acquired noiselet encoded and Fourier encoded data for different acceleration factors (up/down: phase encodes, left/right: frequency encode). \textbf{RF:} shows reference image reconstructed from fully sampled Fourier encoded data; \textbf{RN:} shows reference image reconstructed from fully sampled Noiselet encoded data; \textbf{(a)-(c):} show images reconstructed using Fourier encoding for acceleration factor of 4, 8 and 16 respectively; \textbf{(d)-(f):} show the difference images using Fourier encoding for acceleration factor of 4, 8 and 16 respectively; \textbf{(g)-(i):} show images reconstructed using noiselet encoding for acceleration factor of 4, 8 and 16 respectively; \textbf{(j)-(l):} show the difference images using noiselet encoding for acceleration factor of 4, 8 and 16 respectively. The result here aligns with the simulation results and noiselet encoding outperforms Fourier encoding in preserving resolution. \textbf{(A-H):} Zoomed portion of phantom images reconstructed with Fourier encoding and noiselet encoding with different acceleration factors. \textbf{(A):} shows the original image reconstructed from fully sampled Fourier encoded data; \textbf{(B), (C) and (D):} show the Fourier encoded reconstructed images for acceleration factors of 4, 8 and 16 respectively; \textbf{(E):} shows the image reconstructed from fully sampled noiselet encoded data; \textbf{(F), (G) and (H):} show the noiselet encoded reconstructed images for acceleration factors of 4, 8 and 16 respectively demonstrating that noiselet encoding produces improved resolution images than than Fourier encoding at all acceleration factors.}
\label{APC}
\end{figure}

To validate the feasibility of the proposed reconstruction method, we performed retrospective under-sampling on the acquired noiselet encoded data and Fourier encoded data to simulate accelerated data acquisition. After retrospective under-sampling, the unconstrained optimization program \eqref{eq:min_tv_wav} was solved using the non-linear conjugate gradient method to reconstruct the desired image for different acceleration factors. Fig. \ref{APC} (a)-(c) shows the reconstructed images for the acceleration factors of 4, 8 and 16 on the Fourier encoded data while Fig. \ref{APC} (d)-(f) shows the corresponding difference images. Similarly, Fig. \ref{APC} (g)-(i) shows the reconstructed images for the acceleration factors of 4, 8 and 16 on the noiselet encoded data, and Fig. \ref{APC} (j)-(l) shows the corresponding difference images for noiselet encoded MCS-MRI. These results on the acquired data are consistent with the simulation results and indicate that the noiselet encoding is superior to the Fourier encoding in preserving resolution.

\begin{figure} [ht]
\includegraphics[scale=0.3]{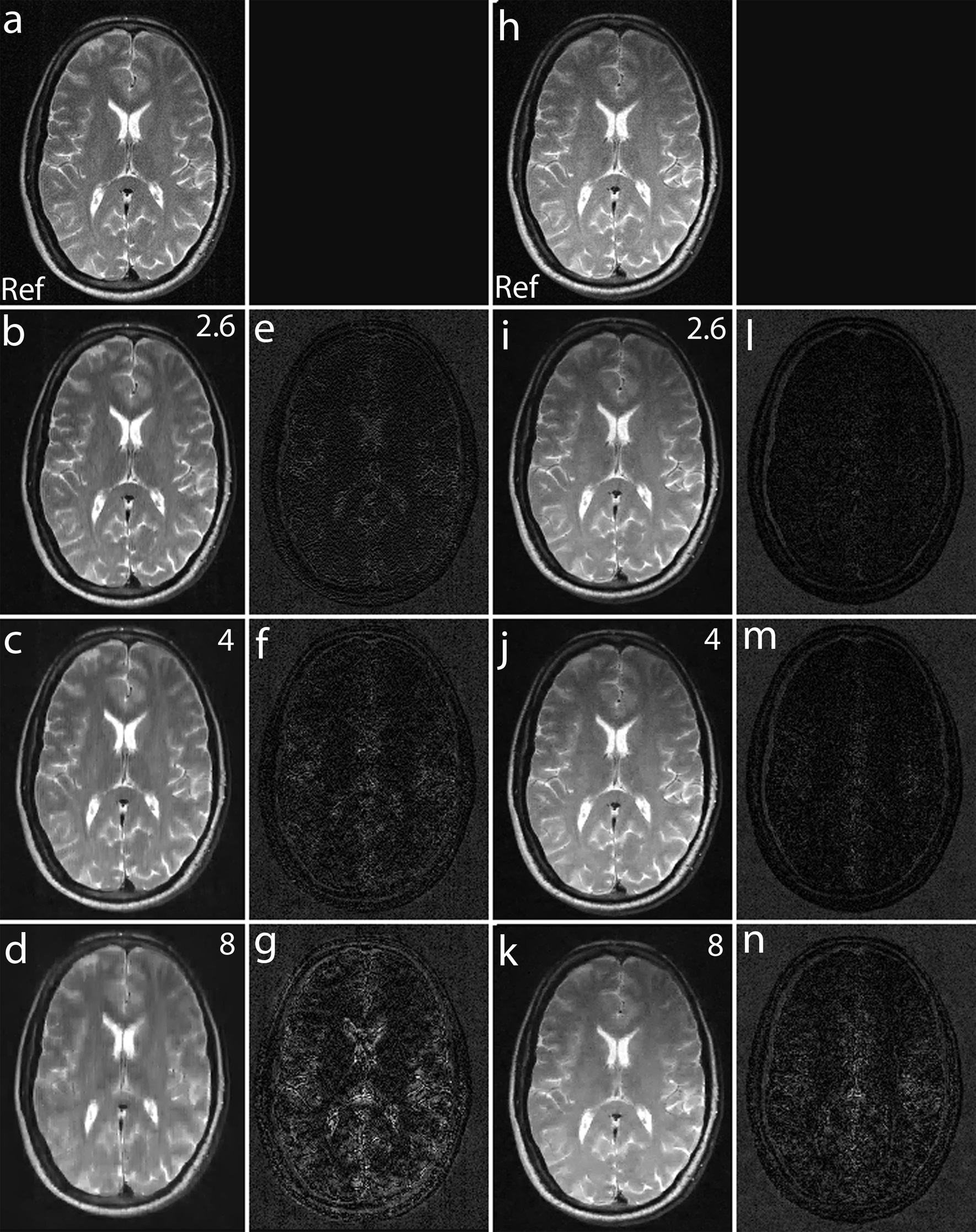}
\centering
\caption{MCS-MRI reconstruction on \textit{in vivo} brain images using acquired noiselet encoded and Fourier encoded data for different acceleration factors (up/down: phase encodes, left/right: frequency encode). \textbf{(a):} shows reference image reconstructed from fully sampled Fourier encoded data; \textbf{(b)-(d):} show images reconstructed using Fourier encoding for acceleration factor of 2.6, 4 and 8 respectively; \textbf{(e)-(g):} show the difference images using Fourier encoding for acceleration factor of  2.6, 4 and 8 respectively; \textbf{(h):} shows reference image reconstructed from fully sampled Noiselet encoded data; \textbf{(i)-(k):} show images reconstructed using noiselet encoding for acceleration factor of 2.6, 4 and 8 respectively; \textbf{(l)-(n):} show the difference images using noiselet encoding for acceleration factor of 2.6, 4 and 8 respectively. It can be seen from the difference images that noiselet encoding outperforms Fourier encoding on the acquired invivo data. The loss in resolution is clearly visible for Fourier encoding at an acceleration factor of 8.}
\label{inv_cmp}
\end{figure}

Fig. \ref{APC} (A-H) shows the zoomed portion of the reconstructed images with Fourier encoding and noiselet encoding. One can distinguish between the small dots in the zoomed images reconstructed with noiselet encoding while it is difficult to distinguish these dots in the images reconstructed with Fourier encoding. This demonstrates that noiselet encoding is able to preserve resolution better than the Fourier encoding. Fig. \ref{inv_cmp} show the images reconstructed with Fourier encoding and noiselet encoding for various acceleration factors on the data acquired for one axial slice of the brain. Since the SNR of the \textit{in vivo} images is less than in the phantom images, reconstruction is shown only up to an acceleration factor of 8. The difference images demonstrate that noiselet encoding outperforms Fourier encoding for all acceleration factors. In particular, at the acceleration factor of 8 the image reconstructed with Fourier encoded data has significantly poorer resolution compared to the image reconstructed with noiselet encoded data.

\FloatBarrier

\begin{figure} [!ht]
\includegraphics[scale=1.0]{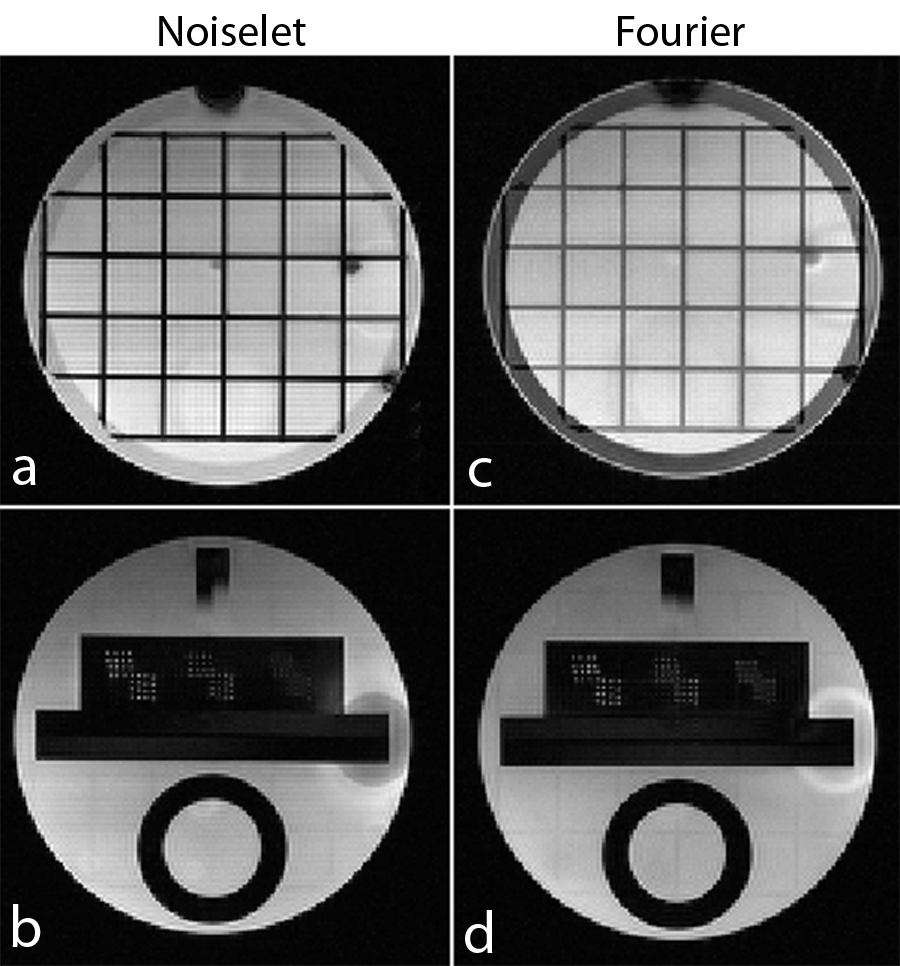}
\centering
\caption{\textcolor[rgb]{0.00,0.00,0.00}{\textbf{(a)} and \textbf{(b):} Two slices of image reconstructed using 3D GRE noiselet encoding sequence, the up/down direction is noiselet encoding direction and left/right direction if Fourier encoding direction; \textbf{(c)} and \textbf{(d):} two slices of image reconstructed with Fourier encoding using 3D GRE sequence.}}
\label{NF_3D}
\end{figure}

\textcolor{black}
{
%\subsection{Noiselet encoding in 3D Gradient Echo Sequence}
Fig. \ref{NF_3D} shows the images reconstructed with noiselet and Fourier encodings using 3D GRE sequence. In our implementation of noiselet encoding in 2D spin echo sequence, the flip angle of 10$^\circ$ was used, which results in loss of some available SNR. Therefore, we have implemented noiselet encoding in 3D Gradient Echo (GRE) sequence as shown in Fig. \ref{RF} (b), which uses noiselets encoding in one direction and Fourier encoding in other two directions. The noiselet encoding is performed using specially selective RF excitation pulse, while the Fourier encoding is performed using gradients. Phantom data was acquired using this sequence with the following parameters: FOV = 200$\times$200 mm$^2$, TE/TR = 6.5/13 ms, slices = 32, FOV of slice = 160 mm, flip angle = 5$^\circ$ and (noiselets) phase encodes  = 256. A 3D Fourier encoded data was also acquired with exactly same parameters. It can be seen from the images that the noiselet encoding provides similar quality of images to that of Fourier encoding. These images are only shown here to demonstrate the feasibility that noiselet encoding can be implemented in 3D GRE sequence.
}

\section{Discussion}
We have presented a new method of encoding the MR data in PE direction with noiselet basis functions for accelerating MRI scans using MCS-MRI reconstruction. The simulation results demonstrate that the proposed encoding \textcolor[rgb]{0.00,0.00,0.00}{gives rise to a multichannel measurement matrix with improved RIP}, and the reconstruction method using the noiselet bases outperforms the conventional Fourier encoding scheme. The mean relative error for noiselet encoding at the acceleration factor of 16 is comparable to that of Fourier encoding  at the acceleration factor of 8, demonstrating that higher acceleration factors can be achieved with noiselet encoding than the Fourier encoding in the MCS framework.

The reconstruction from the noiselet encoding scheme preserves image spatial resolution far better than the Fourier encoding scheme. The Fourier encoding scheme intelligently exploits the property of k-space since most of the energy is concentrated at the center of the k-space. Therefore, densely sampling the center and randomly under-sampling the outer regions of the k-space retains most of the energy in the acquired data. However retaining most of the energy does not imply that most of the information is captured in the acquisition. The low energy (high frequency) component in the outer k-space contains the information about the fine features of the image that the variable density random under-sampling pattern fails to capture. Therefore, the images reconstructed with the Fourier encoding scheme look visually good but have reduced resolution due to insufficient information about the high frequency components in the acquired data. On the other hand, noiselet encoding completely spreads out the energy of the signal in the measurement domain. Therefore each measurement in the noiselet domain has sufficient information to reconstruct the fine details of the image, thus preserving the resolution better than the Fourier encoding.

Noiselet basis functions have some interesting properties that can be exploited, for example noiselets are unitary basis functions and have complex conjugate symmetry. This conjugate symmetry property can be exploited in a way similar to that of the Fourier encoding for partial acquisition \cite{margosian1996partial,mcgibney1993quantitative}. Another interesting property of noiselet basis functions, as for the Fourier basis functions, is that regular under-sampling in the noiselet domain results in aliasing in the image domain. Therefore in the case of regular under-sampling in the noiselet domain, unaliasing with SENSE \cite{pruessmann1999sense} alone can also be used for noiselet encoding.

\textcolor{black}
{
In general the implementation of non-Fourier encoding suffers from a few limitations, and hence the current implementation of noiselet encoding also suffers from these limitations as summarized below. (i) In 2D imaging implementation of noiselet encoding, the excitation of noiselet profile is not slice selective, thus a slice selective 180$^\circ$ pulse is required after excitation, which limits the noiselet encoding to spin echo type sequences. The spin echo sequence always have long TR, therefore the proposed noiselet encoding can only be used for applications requiring long TR, such as structural scans, but will be of little use in dynamic imaging. (ii) In the current implementation, to simplify the design of noiselet excitation pulse we have used direct Fourier transform method that limits the excitation to the low flip angle regime, resulting in the sacrifice of some available SNR. (iii) The duration of noiselet RF excitation pulse is long compared to the conventional sinc RF pulse and the noiselet encoding can only be implemented in one direction in the current implementation. (iv) Due to dielectric effects, etc, in practice, B1 field is always not perfectly homogeneous. Because B1 is used for spatial encoding in the proposed noiselet encoding scheme, B1 inhomogeneity may introduce some perturbations to the noiselet measurement matrix, which in turn may result in some image artifact if the perturbation is large.}

\textcolor{black}
{
The above limitations are not specific to noiselet encoding scheme but are common to all non-Fourier encoding schemes. Here we discuss some probable solutions to the above mentioned problems. (i) The problem of slice selection can be alleviated using a 3D gradient echo (GRE) sequences where a 3D volume can be excited with a noiselet profile in one dimension and other two dimensions can be Fourier encoded. A demonstrative pulse sequence for this solution has been given in Section III and its scan result has been given in Section V,  showing the feasibility of this solution.  (ii) The low tip angle in current implementation is the limitation of the direct Fourier transform method used to compute the RF pulse. It is not an intrinsic limitation of noiselet encoding. Although difficult, computation of large tip angle noiselet RF pulse is possible by using nonlinear computation methods such as direct iterative solution of Bloch equation \cite{xu2008designing} and the SLR method\cite{pauly1991parameter,barralslr}, which are currently being investigated. (iii) The duration of the RF pulse can be reduced by using parallel-transmit and multiple dimensional excitation of noiselet profiles to achieve encoding \cite{grissom2009fast,xu2008designing}. This is our ongoing research.  (iv) The perturbations to the measurement matrix induce an equivalent deterministic noise additive to the measured MR signals. When the inverse noiselet transform is applied directly to the fully sampled dataset to reconstruct the image, a structured artifact may show up if the perturbation is large. This problem can be alleviated when the CS reconstruction method as given in  (\ref{eq:min_tv_wav}) is used for image reconstruction. This is because the CS reconstruction algorithm has inherent denoising capability, which can suppress small perturbations by enforcing the prescribed bound $\epsilon$ on the reconstruction error. See Section II A and the references therein for detailed discussions. For this reason and also because of the high quality of the new 3T scanner used in our experiments, we have not observed structured image artefacts in the experiments presented in Section V.
}

\section{Conclusion}
In this paper we have introduced a method of acquiring data in the noiselet domain and presented a method for the design and implementation of pulse sequences to acquire data in the noiselet domain. The performance of the noiselet encoding has been thoroughly evaluated by extensive numerical analysis, simulation and experiments. The results indicate that \textcolor[rgb]{0.00,0.00,0.00}{the multichannel noiselet measurement matrix has better RIP than that of its Fourier counterpart,} and that the noiselet encoding scheme in MCS-MRI outperforms the conventional Fourier encoding in preserving image resolution, and can achieve higher acceleration factors than the conventional Fourier encoding scheme. The implementation of noiselet encoding by \textcolor[rgb]{0.00,0.00,0.00}{tailoring spin echo and gradient echo sequences} demonstrates that the proposed encoding scheme is pragmatic. The proposed technique has the potential to accelerate image acquisition in applications that require high resolution images.

\bibliographystyle{ieeetr}
\bibliography{references}

\end{document}